\documentclass{aa}  
\usepackage{graphicx,psfig,longtable,lscape,amssym,natbib}  
\newcommand{\ltsima} {$\; \buildrel < \over \sim \;$}  
\newcommand{\gtsima} {$\; \buildrel > \over \sim \;$}  
\newcommand{\lta} {\lower.5ex\hbox{\ltsima}}  
\newcommand{\gta} {\lower.5ex\hbox{\gtsima}}  
\newcommand{\Ha} {H$\alpha$}  
\newcommand{\Hb} {H$\beta$}

\newcommand{\ergscm}{\>{\rm erg}\,{\rm s}^{-1}\,{\rm cm}^{-2}}

\defcitealias{kewley06b}{K06}
\defcitealias{ghisellini01b}{GC01}

\begin{document}

\title{An optical spectroscopic survey of the 3CR sample of radio galaxies
  with $z<0.3$. II.
Spectroscopic classes and accretion modes in radio-loud AGN.\thanks{Based on observations made
    with the Italian Telescopio Nazionale Galileo operated on the island of La
    Palma by the Centro Galileo Galilei of INAF (Istituto Nazionale di
    Astrofisica) at the Spanish Observatorio del Roque del los Muchachos of
    the Instituto de Astrofısica de Canarias.}}
  
\titlerunning{Spectroscopic classes in radio-loud AGN} 
\authorrunning{S. Buttiglione et al.}
  
\author{Sara Buttiglione \inst{1} \and Alessandro Capetti \inst{2} \and
  Annalisa Celotti \inst{1} \and David J. Axon \inst{3,4} \and Marco Chiaberge
  \inst{5,6} \and F. Duccio Macchetto \inst{5} \and William B. Sparks
  \inst{5}}
   
\offprints{S. Buttiglione}
     
\institute{SISSA-ISAS, Via Beirut 2-4, I-34151 Trieste, Italy\\
  \email{buttigli@sissa.it} \and INAF - Osservatorio Astronomico di Torino,
  Strada Osservatorio 20, I-10025 Pino Torinese, Italy \and Department of
  Physics, Rochester Institute of Technology, 85 Lomb Memorial Drive,
  Rochester, NY 14623, USA  \and
School of Mathematical and Physical Sciences, University of Sussex,
Falmer, Brighton BN1 9RH, UK \and Space Telescope Science Institute, 3700 San
  Martin Drive, Baltimore, MD 21218, U.S.A.\and INAF-Istituto di Radio
  Astronomia, via P. Gobetti 101, I-40129 Bologna, Italy}  \date{}

\abstract{In a previous paper we presented a homogeneous and 92 \% complete
  optical spectral dataset of the 3CR radio sources with redshift $<$ 0.3.
  Here we use the emission line measurements to explore the spectroscopic
  properties of the sample. The 3CR sources show a bimodal distribution of
  Excitation Index, a new spectroscopic indicator that measures the relative
  intensity of low and high excitation lines. This unveils the presence of two
  main sub-populations of radio-loud AGN to which we refer to, following
  previous studies, as High and Low Excitation Galaxies (HEG and LEG,
  respectively).  In addition to the two main classes, we find one source with
  a spectrum typical of star forming galaxies, and 3 objects of extremely low
  level of excitation.

  All broad-line objects are HEG from the point of view of their narrow
  emission line ratios and all HEG are FR~II radio-galaxies with log $L_{178}
  $ [erg s$^{-1}$] $ \gtrsim 32.8$.  Conversely LEG cover the whole range of
  radio power encompassed by this 3CR subsample (30.7 $\lesssim$ log $L_{178}
  \lesssim$ 35.4) and they are of both FR~I and FR~II type. The brightest LEG
  are all FR~II. HEG and LEG obey to two (quasi) linear correlations between
  the optical line and extended radio luminosities, with HEG being brighter
  than LEG in the [O~III] line by a factor of $\sim$ 10. HEG and LEG are
  offset also in a plane that compares the black hole mass and the ionizing
  nuclear luminosity.

  However, although HEG are associated with higher nuclear luminosities, we
  find LEG among the brightest radio sources of the sample and with a clear
  FR~II morphology, indistinguishable from those seen in HEG.  This suggests
  that LEG are not simply objects with a lower level of accretion.  We
  speculate that the differences between LEG and HEG are related to a
  different mode of accretion: LEG are powered by hot gas, while HEG require
  the presence of cold accreting material. The high temperature of the
  accreting gas in LEG accounts for the lack of ``cold'' structures (i.e.
  molecular torus and Broad Line Region), for the reduced radiative output of
  the accretion disk, and for the lower gas excitation.

   }

\maketitle
  
\section{Introduction}
\label{intro}

Radio galaxies (RG) are an important class of extragalactic objects for many
reasons. 
Studies of radio-loud AGN are the key to understand the
processes leading to the ejection of material in relativistic jets and its
connection with gas accretion onto the central black holes, the way in which
different levels of accretion are related to the process of jets launching,
the origin of the AGN onset, and its lifetime.
But the intense nuclear activity
can also influence the star formation history and the properties of
the ISM and ICM, thus representing a fundamental ingredient for the 
evolution of their hosts and their large scale environment. 

RG have been historically classified according to their radio
morphology, following the \citet{fanaroff74} criteria: a FR~I source has
bright jets rising from the nucleus, while a FR~II has two bright hot
spots far from it. They also noted that FR~II are mostly found at high radio
luminosities, while FR~I are associated to weaker radio sources.  The two FR
classes also differ, at least statistically, from several other points of
view, such as the environment \citep{zirbel97} and host
luminosities \citep{govoni00}. However, it soon became apparent that the
transition between the two classes is continuous and objects of intermediate
radio structure do exist \citep[e.g. ][]{capetti95}. A class of
hybrid double sources, with a FR~I jet on one side and a FR~II lobe on the
other, was also unveiled by \citet{gopal00}. This supports explanations for
the FR dichotomy based upon jet interaction with the external medium, arguing
against interpretations based on intrinsic differences in the central engine.
Furthermore, the multi-wavelength behavior of the nuclear emission in
RG does not appear to be directly related to the differences
between the two FR classes, but it is more closely linked to their
spectroscopic nuclear properties \citep[e.g.][]{chiaberge00,chiaberge:fr2}.

Optical spectroscopic information can clearly play a major role in gaining a
better understanding of the properties of the central engines of RG.
\citet{heckman80} and \citet{baldwin81} proposed to use optical line ratios as
diagnostic tools to classify emission-line objects in general and AGN in
particular.  They introduced diagnostic diagrams comparing selected emission
line ratios, able to distinguish H~II regions ionized by young stars from gas
clouds ionized by nuclear activity. Moreover AGN were separated into Seyferts
and Low Ionization Nuclear Emission-line Regions \citep[LINERs,][]{heckman80}
based on the relative ratios of the optical oxygen lines ([O I]$\lambda$6364,
[O II]$\lambda$3727, and [O III]$\lambda$5007). Subsequently
\citet{veilleux87} revised the definition of the diagnostic diagrams, using
only ratios of lines with small separation in wavelength, thus reducing the
problems related to reddening as well as to uncertainties on the flux
calibration of the spectra.  They used the following line combinations:
[O~III]/H$\beta$ as a function of [N~II]$\lambda$6583/\Ha,
[S~II]$\lambda\lambda$6716,6731/\Ha, and [O~I]/\Ha.  The separation between
AGN and HII regions, initially introduced empirically, was calibrated
theoretically by \citet{kewley01}. More recently, \citet[ hereafter
K06]{kewley06} selected a sample of $\sim 85000$ emission line galaxies from
the SDSS, finding that Seyferts and LINERs form separated branches on the
diagnostic diagrams. They suggested that the observed dichotomy corresponds to
the presence of two sub-populations of AGN associated with different accretion
states.

An attempt to adopt a similar scheme for the optical classification focusing
on radio-loud galaxies was made by \citet{laing94} on a sub-sample of 3CR
radio-galaxies, selected imposing z$<$0.88, V$<$20, and 0$^h <$ RA $< 13^h$.
They put on firmer ground the original suggestion by \citet{hine79} that FR~II
sources can be distinguished into subclasses.  They proposed a separation into
high excitation galaxies (HEG, defined as galaxies with [O~III]$/$\Ha$>0.2$
and equivalent width (EW) of [O III] $>$ 3 \AA) and low excitation galaxies
(LEG).  \citet{tadhunter98} found a similar result from an optical
spectroscopic study of the 2Jy sample, in which a sub-class of Weak-Line Radio
Galaxies (the sources with EW of [O III])$<10$\AA) stands out due to a low
ratio between emission line and radio luminosities as well of [O III]/[O II]
line ratio.

Emission line luminosities show a broad connection with radio power, as
verified by many studies
\citep[e.g.][]{baum89a,baum89b,rawlings89,rawlings91}.  This relation holds
also for compact steep spectrum (CSS) \citep{morganti97} and GHz peaked
spectrum (GPS) sources \citep{labiano08}.  \citet{willott99} demonstrated that
the dominant effect is a strong positive correlation between $L_{\rm radio}$
vs $L_{\rm line}$ and not to a common dependence of these quantities on
redshift. This relationship points to a common energy source for both the
optical line and the radio emission, and suggests the radio and line
luminosities of RG are determined, to first order, by the
properties of their central engines.  \citet{baum89b} noticed that the large
scatter (roughly one order of magnitude) in $L_{\rm radio}$ for a given
$L_{\rm line}$ suggests that other factors do play a secondary role, such as
the environment.  Another possibility is that the line and radio luminosities
may be independently correlated with a third parameter, e.g. the amount of
cold gas present on the kiloparsec scale.  A correlation between emission
lines and core radio powers, although weaker than that observed for $L_{\rm
  radio}$ and $L_{\rm line}$, is also found \citep[e.g.
][]{baum89b,rawlings89,rawlings91} suggesting that the total radio luminosity
is separately correlated with the [O~III] and the core radio power.

\citet{morganti92} extended the line-radio connection towards lower radio
luminosities considering sources from the B2 sample, predominantly FR~I, 
noting a flattening in the correlation. This result is supported also
by the analysis by \citet{zirbel95}, who also report that FR~II sources
produce about 5-30 times more emission line luminosity than FR~I for the same
total radio power \citep[see also][]{tadhunter98,wills04}.

In addition to the diagnostics derived from the narrow lines, optical spectra
also reveal the presence of prominent permitted broad lines in a significant
fraction of RG.  Unifying models for active galaxies ascribe the
lack of broad lines to selective obscuration and differences of orientation
between narrow and broad lined AGN \citep[see ][]{antonucci93,urry95}.

However, despite the massive amount of spectroscopic data collected over the
last decades for RG, there are still several key questions waiting for clear
answers: are there indeed two (or more) distinct populations of RG? On which
basis they can be separated? Which are the physical parameters setting the
division? Which is the relationship between the optical and radio properties?
When are broad emission lines observed?  Which processes set the presence and
detectability of broad lines and how do they fit in the unified models for
active nuclei? How do the spectroscopic properties of radio-loud and
radio-quiet AGN compare?

A necessary ingredient required to enable a detailed exploration of
spectroscopic properties of radio-loud AGN is a dataset as homogeneous and
complete as possible.  This is one of the aims of the optical spectroscopic
survey that we performed with the Telescopio Nazionale Galileo, presented
in \citet[ hereafter Paper I]{buttiglione09}.

We considered the 113 3CR sources at redshift $<$ 0.3, with an effective
coverage of the sub-sample of 92 \% (i.e. 104 sources).  For 18 sources the
spectra are available from the Sloan Digital Sky Survey (SDSS) database
\citep{york00, st02, yip04}, Data Release 4,5,6.
For these sources the observations provide uniform and uninterrupted coverage
of the key spectroscopic optical diagnostics ratios.
The data quality is such that the \Ha\ line is detected
in all but 3 galaxies and in the majority ($\sim 75$\%) of the objects all key
emission lines (i.e.  H$\beta$, [O~III]$\lambda\lambda$ 4959,5007 \AA,
[O~I]$\lambda\lambda$ 6300,64 \AA, H$\alpha$, [N~II]$\lambda\lambda$ 6548,84
\AA, [S~II]$\lambda\lambda$ 6716,31 \AA) 
required to construct diagnostic diagrams are detected.

The attractiveness of the 3CR catalog of radio sources as a basis for such a
study is obvious, being one of the best studied sample of RG. Its
selection criteria are unbiased with respect to optical properties and
orientation, and it spans a relatively wide range in redshift and radio power.
A vast suite of observations is available for this sample, from multi-band HST
imaging to observations with Chandra, Spitzer and the VLA, that can be used
to address the issues listed above in a multiwavelength approach.

The paper is organized as follows: in Sect. \ref{optclas} we derive and
explore the spectroscopic diagnostic diagrams for the 3CR sample of sources
with $z \leq 0.3$. In Sect. \ref{or} we compare optical (line and host)
luminosity and radio emission. Results are discussed in Sect. \ref{discussion}
and summarized in Sect. \ref{fine}, where we also present our conclusions.

Throughout, we adopted $H_o = 71$ km s$^{-1}$ Mpc$^{-1}$, $\Omega_{\Lambda} =
0.73$ and $\Omega_m = 0.27$.

\section{Spectroscopic classification of 3CR sources}
\label{optclas}

\begin{figure*}[htbp]
  \centerline{ 
    \psfig{figure=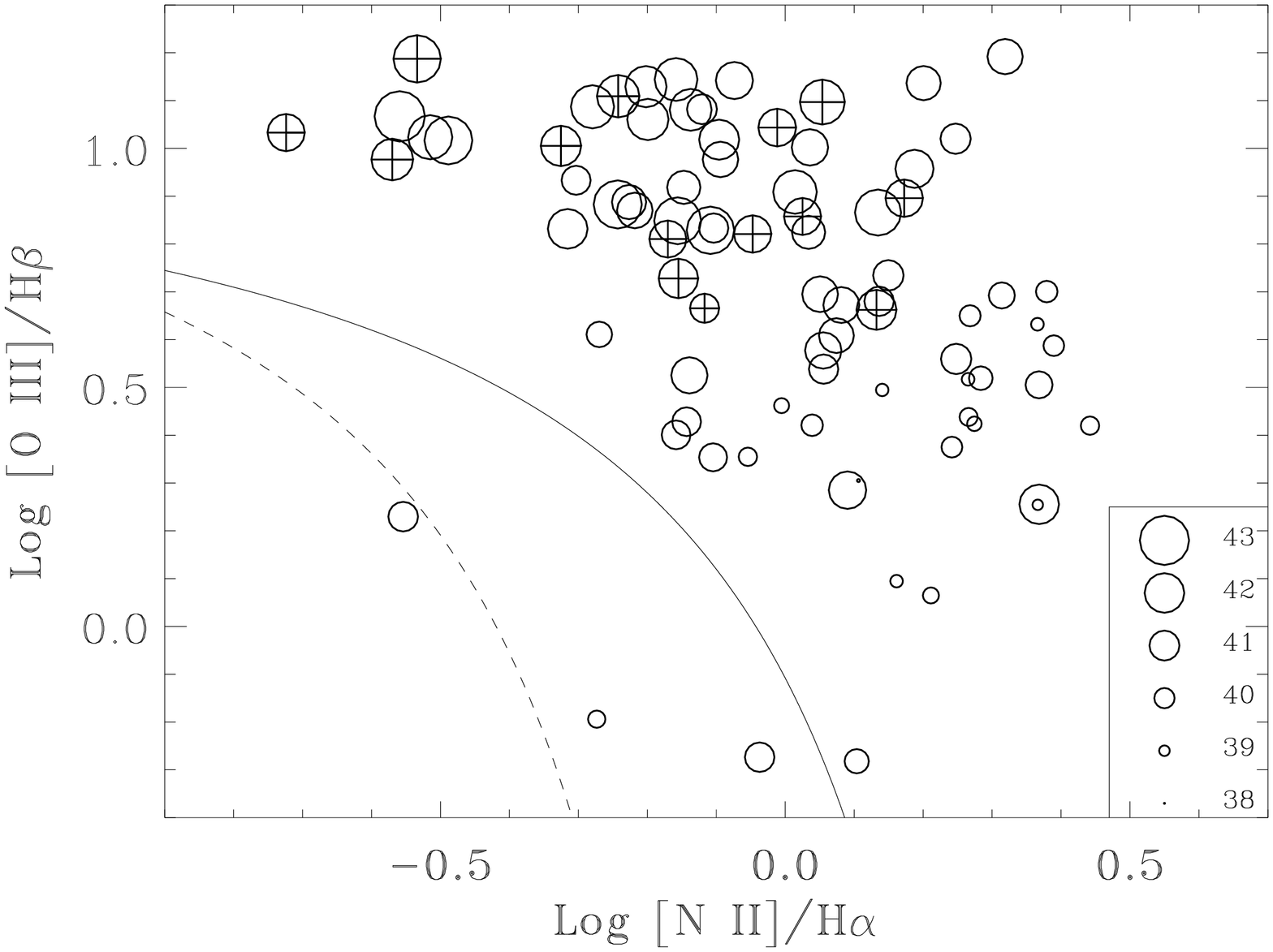,angle=90,width=0.33\linewidth}
    \psfig{figure=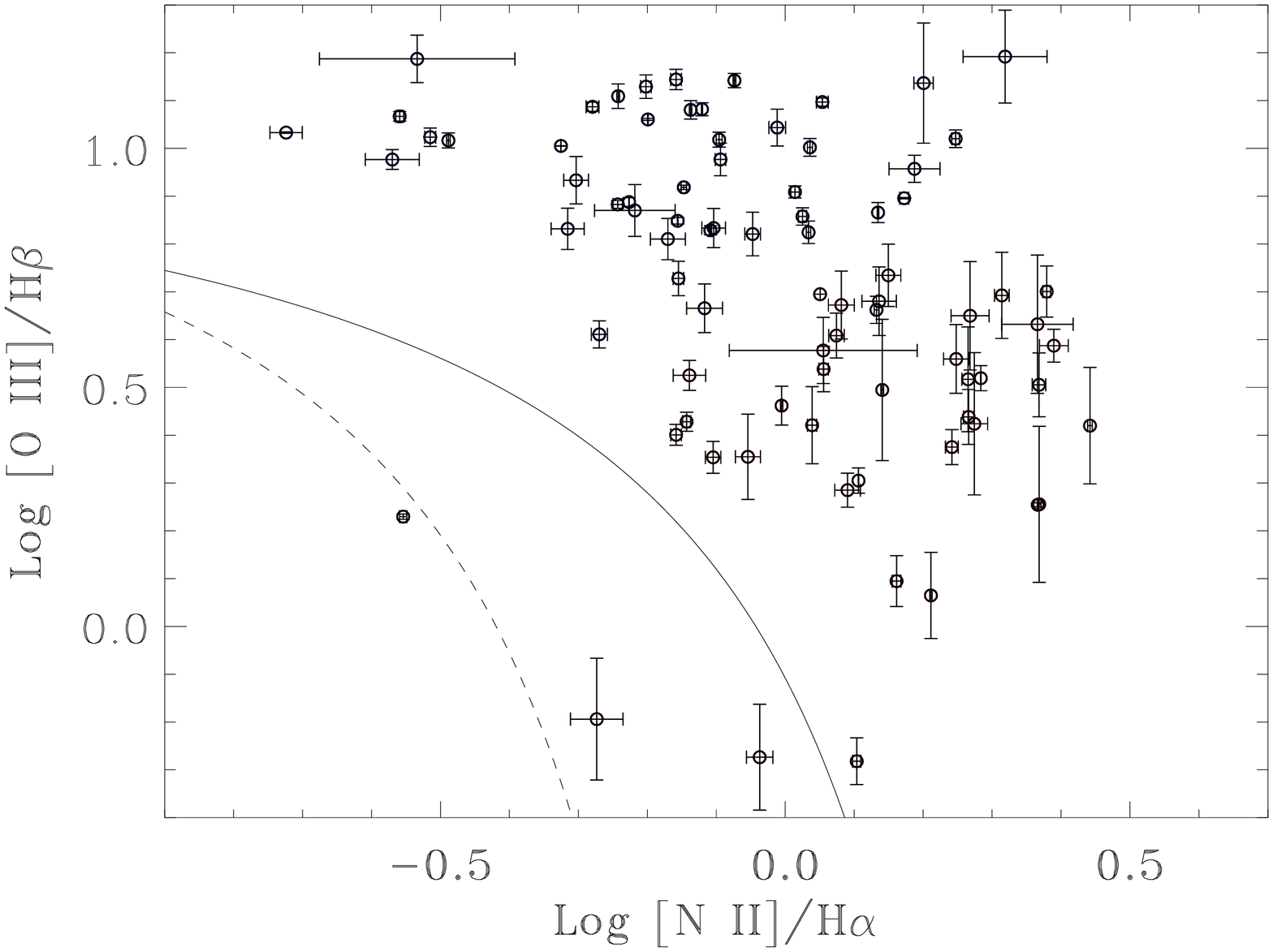,angle=90,width=0.33\linewidth} 
    \psfig{figure=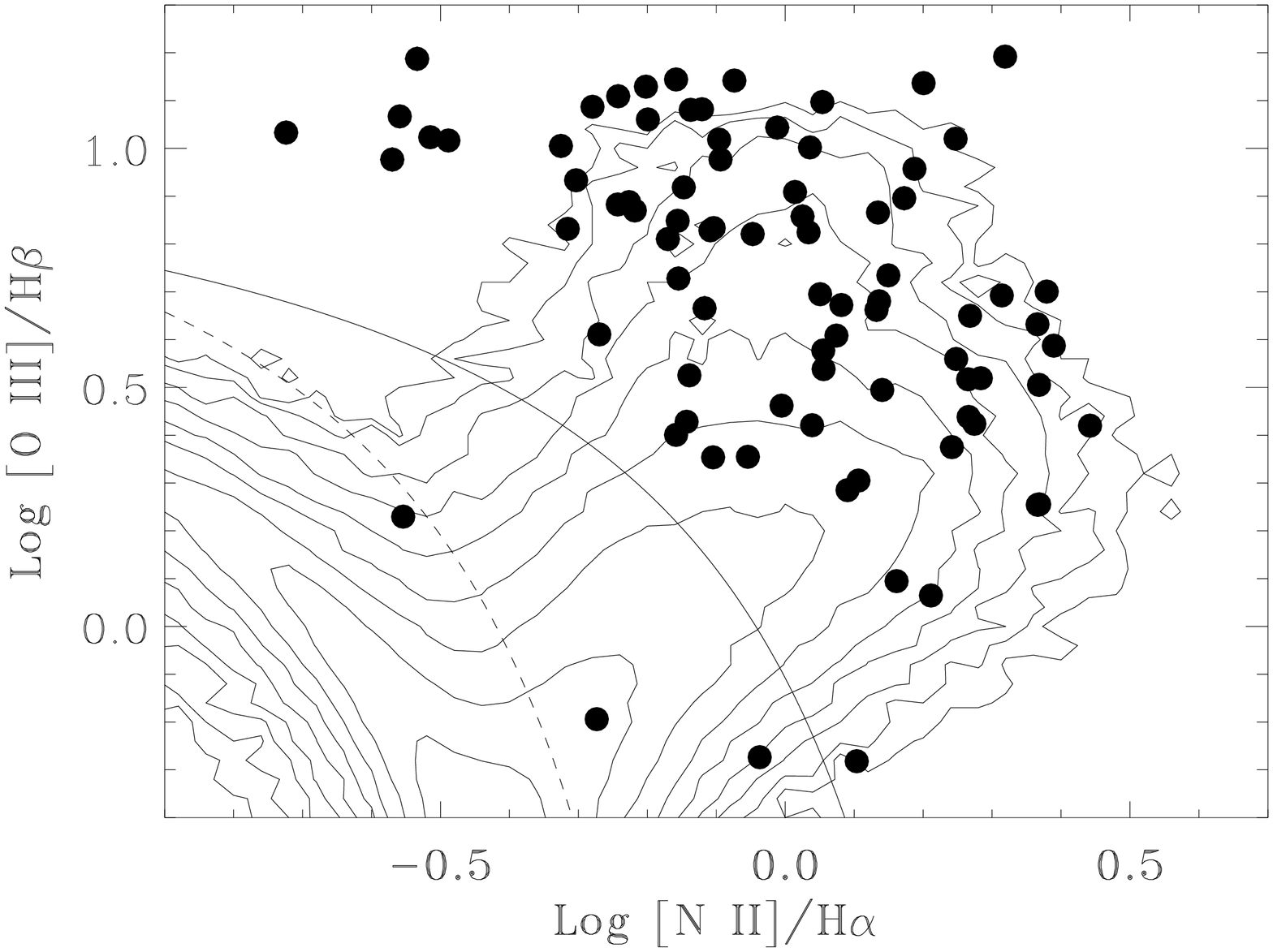,angle=90,width=0.33\linewidth}}
  \centerline{ 
    \psfig{figure=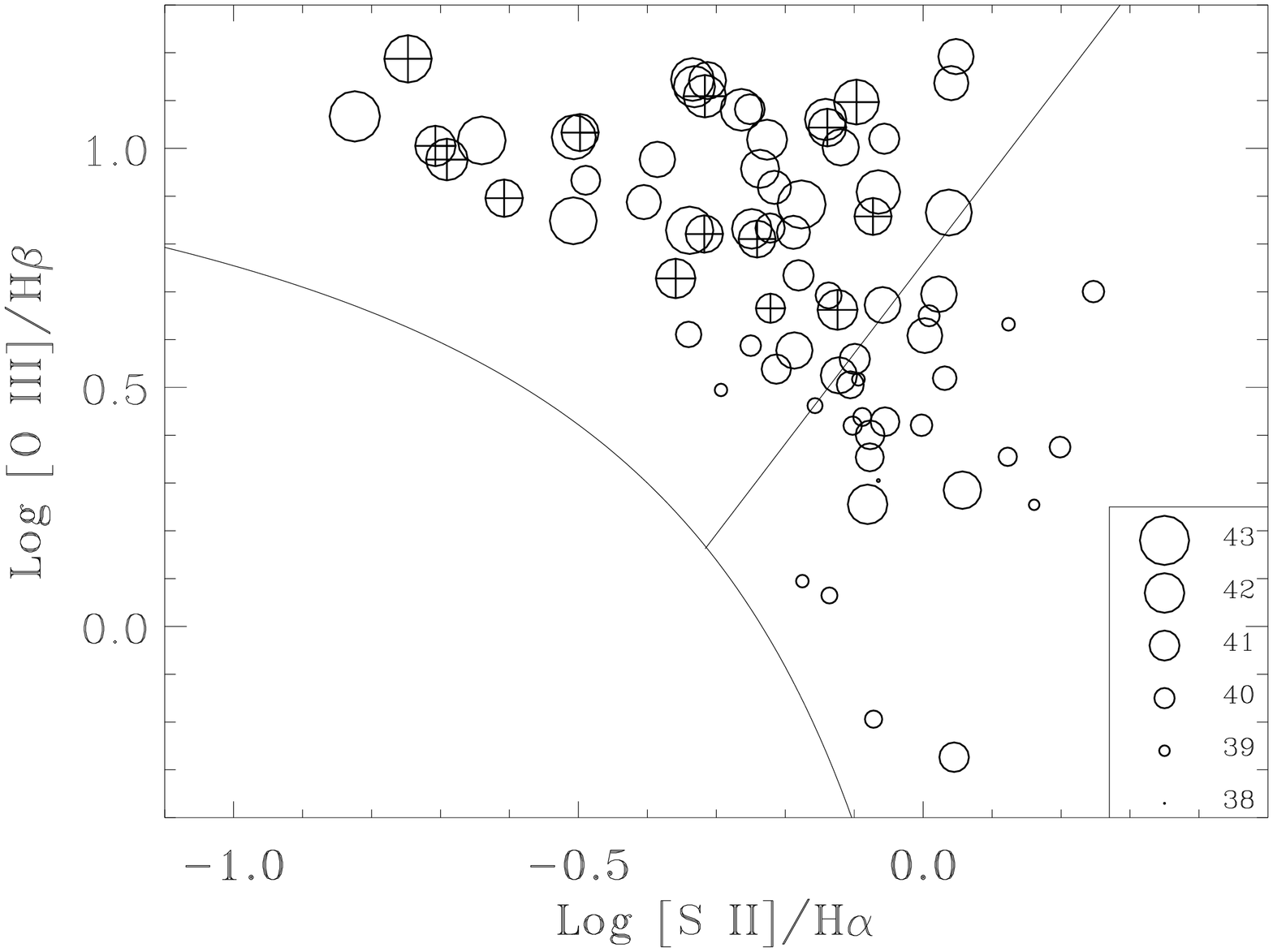,angle=90,width=0.33\linewidth}
    \psfig{figure=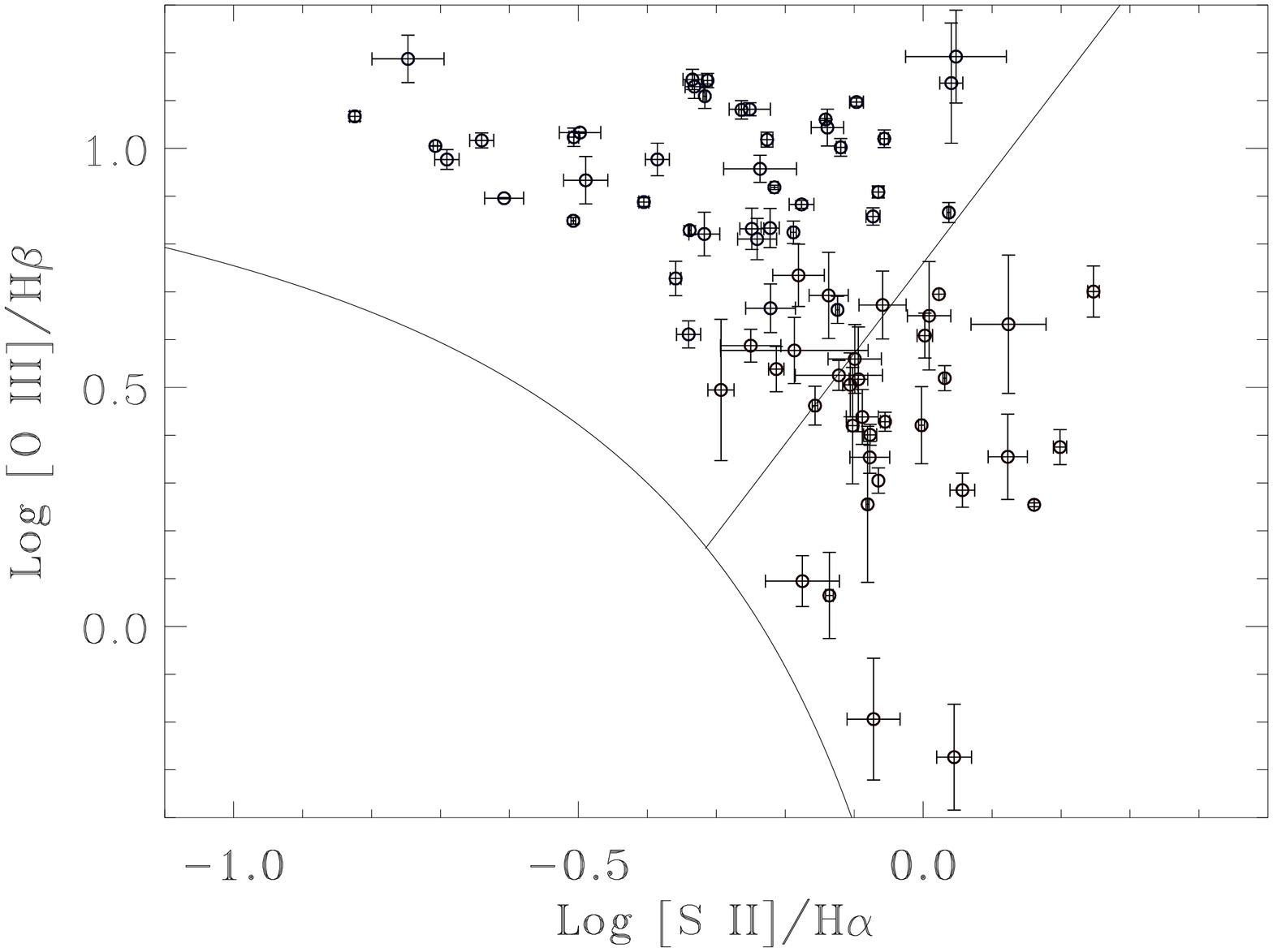,angle=90,width=0.33\linewidth} 
    \psfig{figure=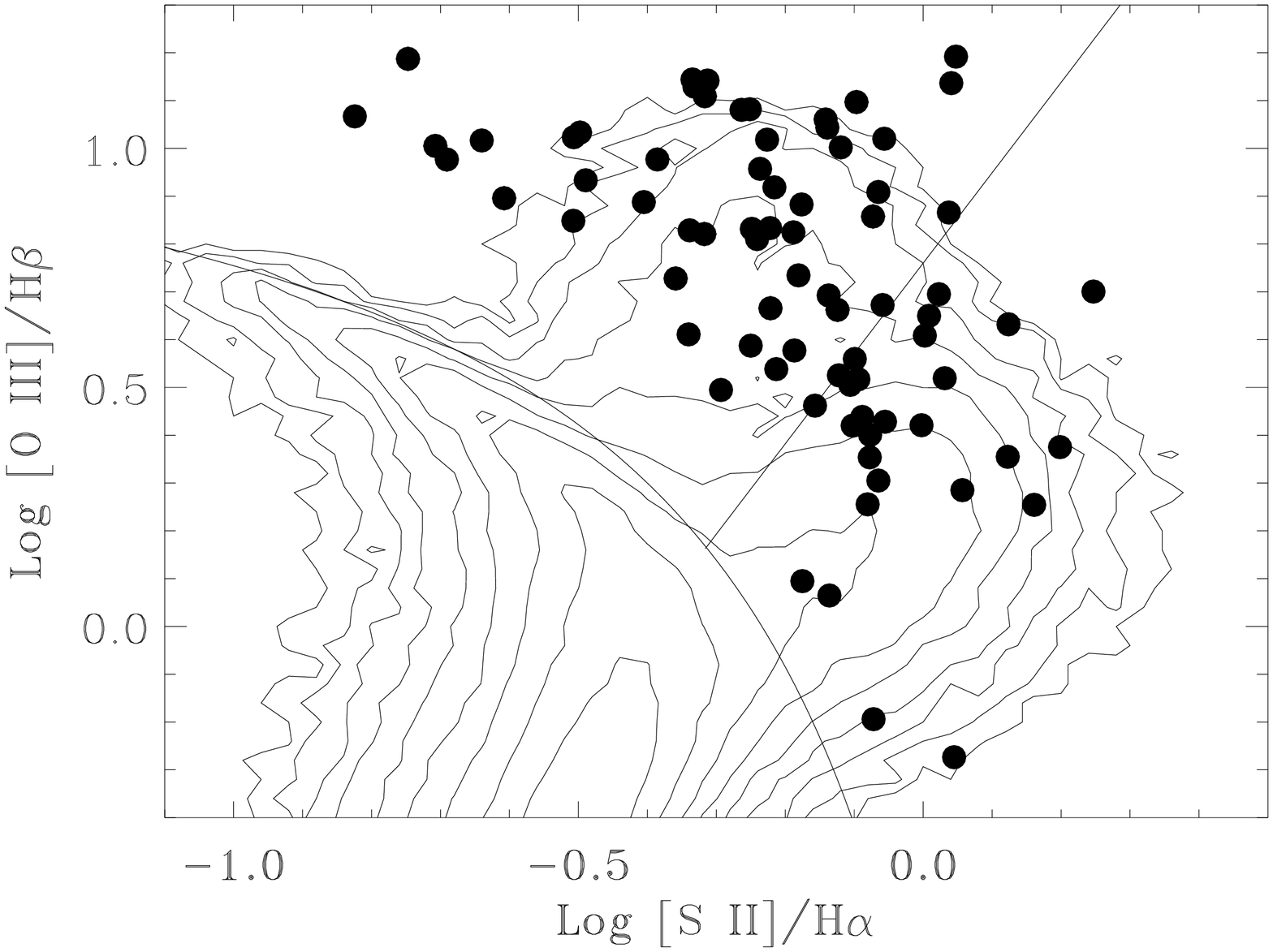,angle=90,width=0.33\linewidth}}
  \centerline{ 
    \psfig{figure=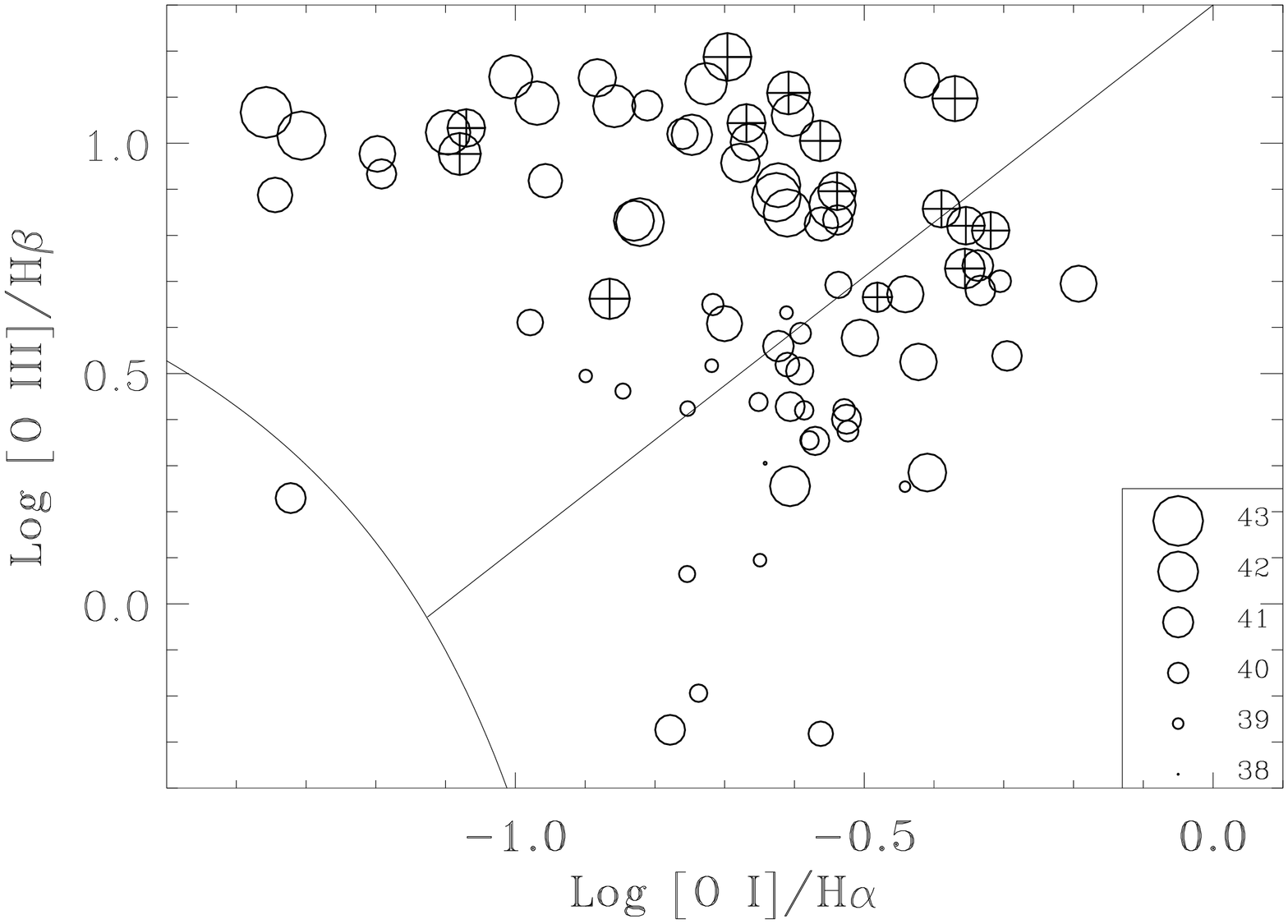,angle=90,width=0.33\linewidth}
    \psfig{figure=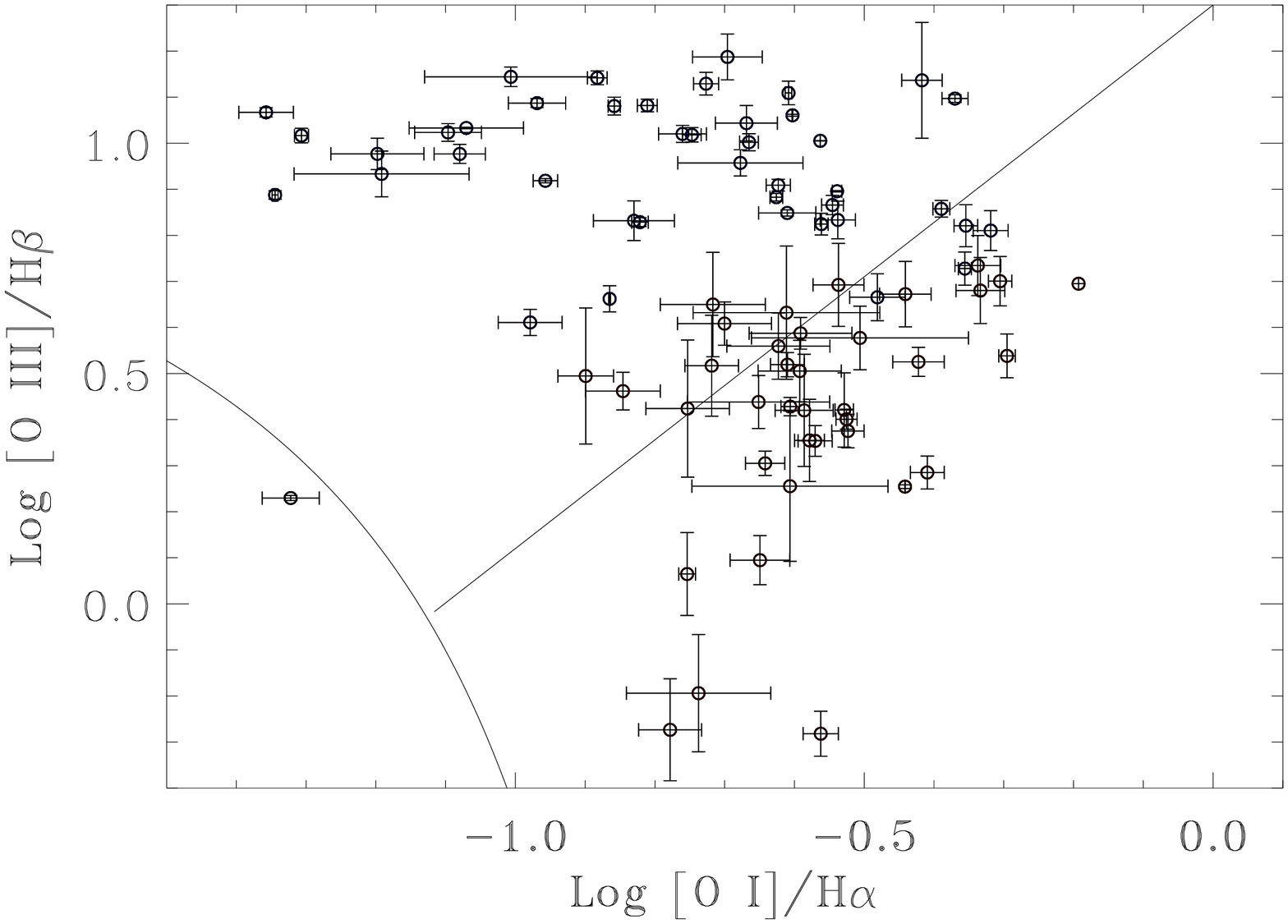,angle=90,width=0.33\linewidth} 
    \psfig{figure=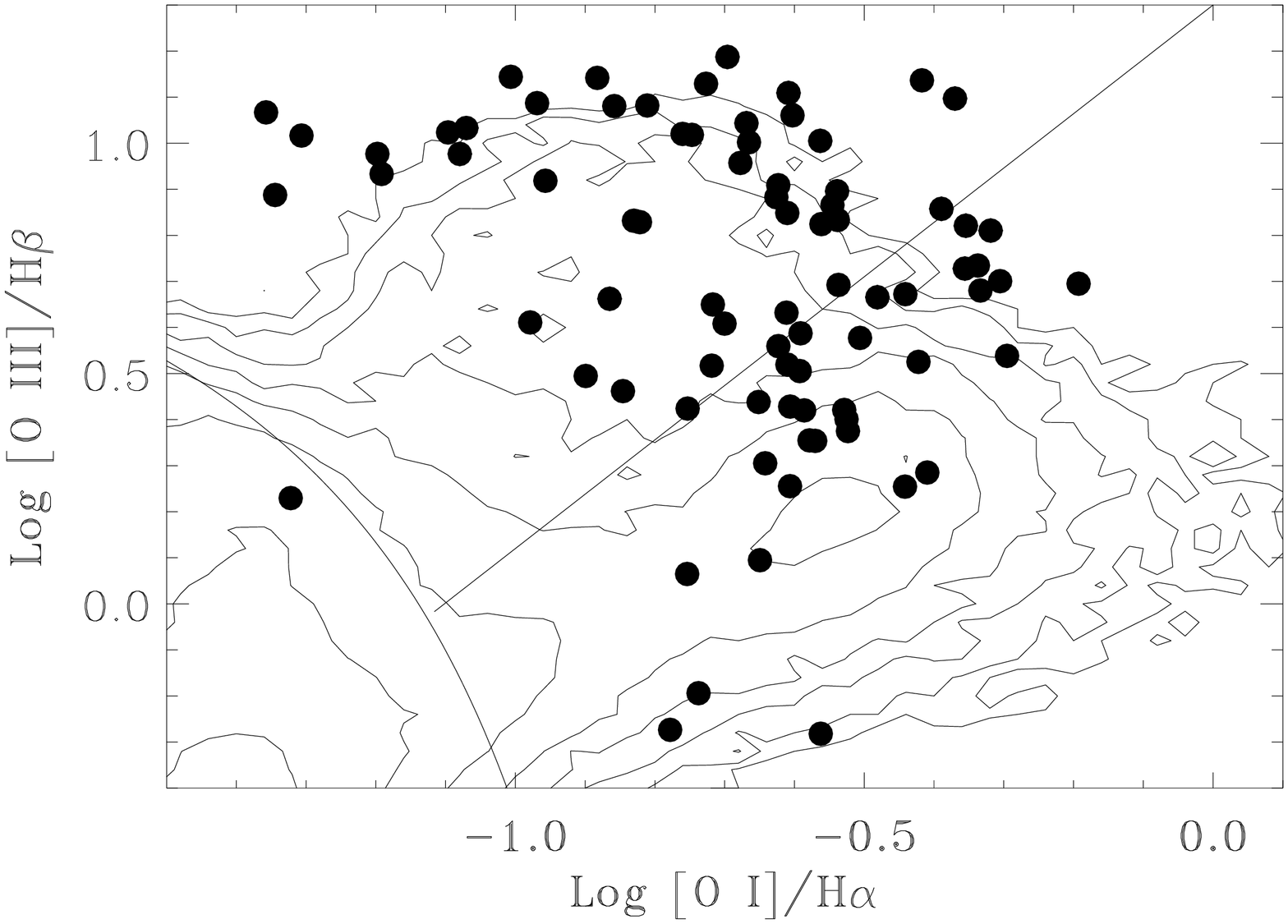,angle=90,width=0.33\linewidth}}
  \caption{\label{dd} Diagnostic diagrams. Left column: 3CR sources with
    symbol sizes increasing with [O~III] luminosity (log $L_{\rm [OIII]}$(erg
    s$^{-1}$)); the galaxies showing a broad \Ha\ component are represented by
    crossed circles. Central column: the same plot with error bars. Right
    column: comparison between 3CR (black dots) and the density of SDSS
    emission line sources (contour lines) from K06. The ordinates are the
    ratio log [O~III]/\Hb.  The first row shows the log [N~II]/\Ha\ ratio, the
    second row the log [S~II]/\Ha\ ratio, while the third row shows log
    [O~I]/\Ha.  The curves divide AGN (above the solid curved line) from
    star-forming galaxies.  Between the dashed and solid curves there are the
    composite galaxies \citepalias{kewley06b}.  In the bottom two rows, the
    straight solid line divides Seyferts (upper left region) from LINERs
    (right region).}
\end{figure*}

\subsection{Spectroscopic diagnostic diagrams}
\label{3cdiagr}

We used the emission lines intensities measured in Paper I\footnote{We removed
3C~270 from the list of observed objects since the
SDSS fiber was not positioned on the galaxy's nucleus (Christian Leipski,
private communication).} to explore the
spectral properties of the 3CR sample taking advantage of the spectroscopic
diagnostic diagrams \citep{kewley06}. More specifically we constructed the
diagrams that compare the [O~III]$\lambda$5007$/$H$\beta$ ratio with
[N~II]$\lambda$6583$/$H$\alpha$, [S~II]$\lambda\lambda$6716,6731$/$H$\alpha$
and [O~I]$\lambda$6364$/$H$\alpha$ shown in Fig. \ref{dd}.
Ratios involving line upper limits are not considered. 

In the left hand panels of this figure, the 3CR sources are indicated by
circles with sizes proportional to the [O~III] luminosity, in the central
panels the errors on the line ratios are shown and in the right panels the 3CR
sources are compared with the SDSS emission line galaxies from
\citetalias{kewley06b}.  The solid lines divide sources into star-forming
galaxies (lower left region of the diagram), Seyferts (top left region) and
LINERs (bottom right region) according to \citetalias{kewley06b}.

In the first diagnostic diagram (first row) of Fig. \ref{dd}, log
[O~III]$/$H$\beta$ versus log [N~II]$/$H$\alpha$, all 3CR sources are located
in the AGN region with only a few exceptions: one object (3C~198\footnote{The
  source is not present in the middle diagram because the [S~II] lines are not
  covered by its SDSS spectrum.}) falls among the star-forming galaxies. Two
objects (3C~028 and 3C~314.1) are located in the composite region and,
together with 3C~348, have extremely low [O~III]$/$H$\beta$ ratio ($\sim
0.5$); however, they are well into the AGN region in the other two
  diagrams.  The location of the remaining sources appears to be related to
their line luminosity. Powerful sources ($L_{\rm [OIII]}\gtrsim10^{41.5}$ erg
s$^{-1}$) are mostly found along a horizontal strip around log[O~III]$/$\Hb
$\sim1$. We also note that all 3CR galaxies in which we detected a broad \Ha\
component belong to this sub-group.  In contrast, fainter sources ($L_{\rm
  [OIII]}\lesssim10^{41.5}$ erg s$^{-1}$) are instead generally distributed in
a region around log[O~III]$/$\Hb $\sim0.5$.  The sample divides roughly
equally into powerful and faint sources.  The horizontal spread is
significantly broader for the bright galaxies as they extend to lower values
of the [N~II]/H$\alpha$ ratio.

In the second and third diagnostic diagrams, log [O~III]$/$H$\beta$ versus log
[S~II]$/$\Ha\ \footnote{For 3 sources (namely 3C~028, 3C~035, and
    3C~066B) we could only measure the [S II]$\lambda$6716 line. We estimated
  the total [S II] flux adopting equal intensities in the two lines and
  increased the line error to consider this approximation.}  and log
[O~I]$/$\Ha, the sources show very similar distributions to that seen in the
previous diagram.  In these cases, we also take advantage of the separation
into Seyfert and LINERs (i.e. into high and low ionization galaxies) proposed
by \citetalias{kewley06b} and graphically represented by the oblique solid
lines.  All powerful 3CR are located above the separation. Conversely, fainter
sources straddle the Seyferts/LINERs divide.  In particular, considering the
log [O~I]$/$H$\alpha$ ratio, they cover a region elongated along the
Seyferts/LINERs separation line.

\begin{figure}[htbp]
  \centerline{ 
    \psfig{figure=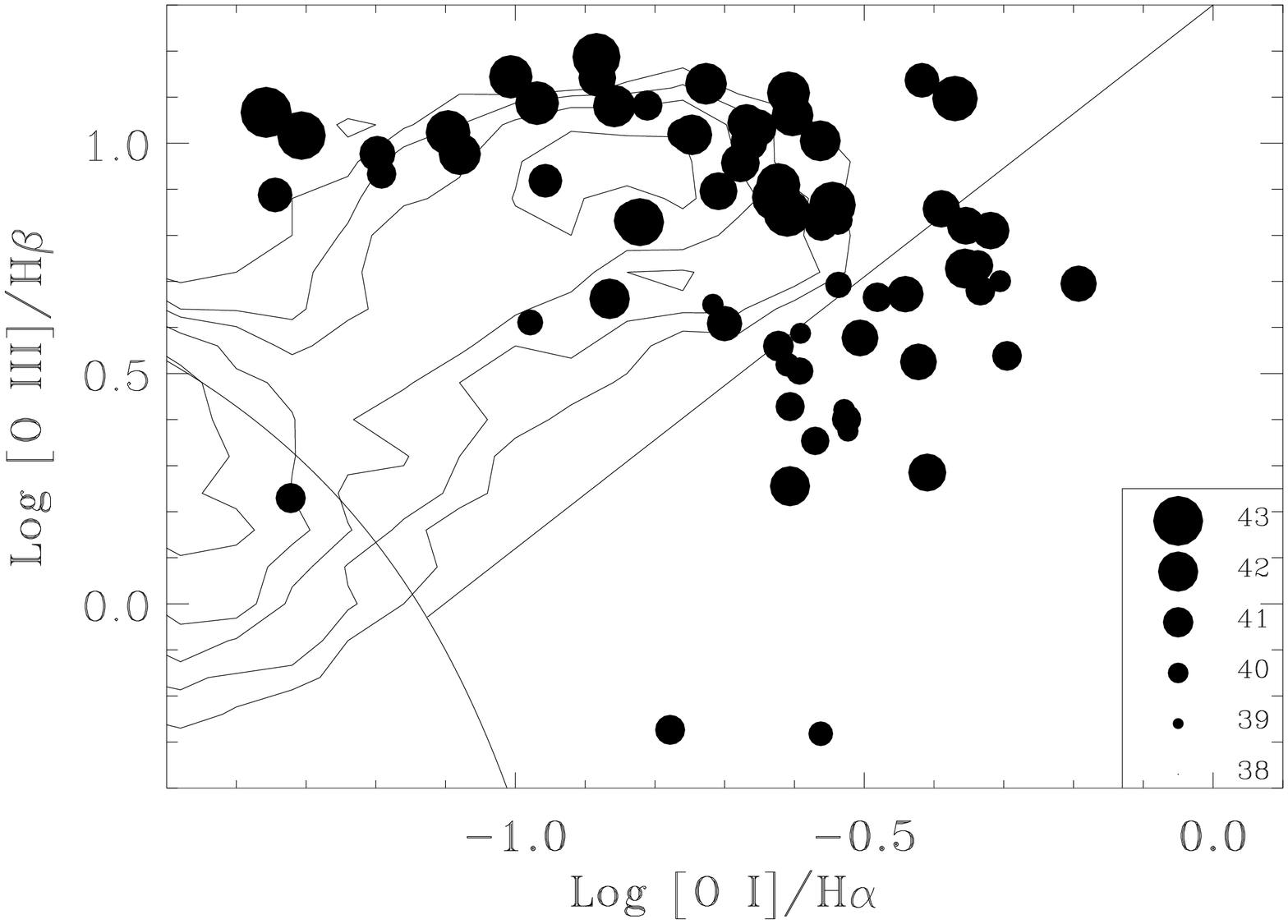,angle=90,width=0.99\linewidth}}
  \caption{\label{sdss4045}  log [O~III]$/$H$\beta$ versus log
    [O~I]$/$H$\alpha$ diagram only for sources
    with $L_{\rm [OIII]}>10^{40}$ erg s$^{-1}$: 
contour lines represent the SDSS sample while
    black circles indicate the 3CR sources. The symbol size increases with
    [O~III] luminosity. As in Fig. \ref{dd}, the curved line divides the
    star-forming galaxies and AGN, while the oblique line divides Seyferts and
    LINERs (according to \citetalias{kewley06b}).  
  }
\end{figure}

In the right column of Fig. \ref{dd} our data (circles) are compared with the
SDSS data from \citetalias{kewley06b} (contour lines): SDSS data are composed
of $\sim$ 85000 galaxies, mainly radio quiet AGN (Seyferts, LINERs) and
star-bursts galaxies. As already mentioned most 3CR sources fall in the AGN
region, but several differences can be seen comparing 3CR and SDSS AGN.  First
of all, a group of powerful 3CR sources cover a region with low values of
[N~II]$/$\Ha\ and [S~II]$/$\Ha\ (top left of the diagrams): this group of
sources seems to have no corresponding sources among SDSS objects.  In the
[O~I]$/$H$\alpha$ diagram, 3CR sources are mainly distributed along the edge
of the region covered by SDSS galaxies, both Seyfert and LINERs. In the other
two diagrams the situation is less extreme, but many 3CR sources are located
in areas of low AGN density.

These differences are probably not so surprising, considering the strong
mismatch in line luminosity between the two samples. In fact, the mean [O~III]
luminosity of the SDSS sample is about 30 times lower than the 3CR sample.
Indeed, the majority of LINERs has luminosities in the range $10^{38}\lesssim L_{\rm
  [OIII]}\lesssim 10^{39}$ erg s$^{-1}$, the SDSS Seyferts are mostly in the range
$10^{39}\lesssim L_{\rm [OIII]}\lesssim 10^{41}$ erg s$^{-1}$ while half of the 3CR sources
have luminosities in excess of $L_{\rm [OIII]}\sim10^{41}$ erg s$^{-1}$.  The
luminosity mismatch is well visible in Fig.  \ref{sdss4045} where only sources
with $L_{\rm [OIII]}\gtrsim10^{40}$ erg s$^{-1}$ are plotted: among the SDSS
sources, LINERs are substantially absent at these luminosities, and only a well
defined Seyfert `finger' is present. Conversely, this threshold
selects most 3CR galaxies, including those with low [O~III]/\Hb\ ratios.
 
An important consequence is that it is probably inappropriate to
blindly apply the Seyferts/LINERs separation found by \citetalias{kewley06b} 
to the 3CR sources. 
Similarly, we cannot guarantee that two sub-populations of radio-loud
AGN exist based on the results found for (mostly) radio-quiet AGN of much
lower activity level.
 
\subsection{A new spectroscopic classification scheme}
\label{hegleg}

In addition to the points discussed in the previous section there are 
two further complications in the attempt to understand the spectroscopic
properties of the 3CR radio-galaxies. 

The first is related to the relatively large fraction of SDSS galaxies whose
location with respect to the curves separating the various classes of emission
line galaxies vary in the different diagnostic diagrams, defined by
\citetalias{kewley06b} as `ambiguous galaxies'.  While this does not represent
a significant problem for the interpretation of the properties of the SDSS
sources, thanks to the extremely large number of objects, this effect can have
a strong impact due to the much smaller size of the 3CR sample. We decided to
go beyond this complication by introducing new spectroscopic indicators.  We
estimate for all narrow line objects the average of the low ionization
lines ratios, i.e.  1/3 (log [N~II]$/$\Ha + log [S~II]$/$\Ha + log
[O~I]$/$\Ha) that we define as Line Ratios Index (LRI) as well as the
Excitation Index (E.I.), log [O~III]/\Hb\ - LRI. These indices are clearly more
stable than the single line ratios. In particular, the Excitation Index
represents the overall ratio of high and low excitation emission lines in each
source.

The second issue is the presence of a substantial group of 3CR galaxies with
strong broad emission lines. In these objects, the measurement of the narrow
emission lines is less reliable due to the presence of the strong nuclear
continuum and to the complexity of the profiles of the broad lines. In
particular, as explained in Paper I, we expect a general
over-estimate of the \Hb\ line luminosity. For this reason we defer the
discussion of broad lined objects to Sect. \ref {blo}.

\begin{figure}[htbp]
  \centerline{ \psfig{figure=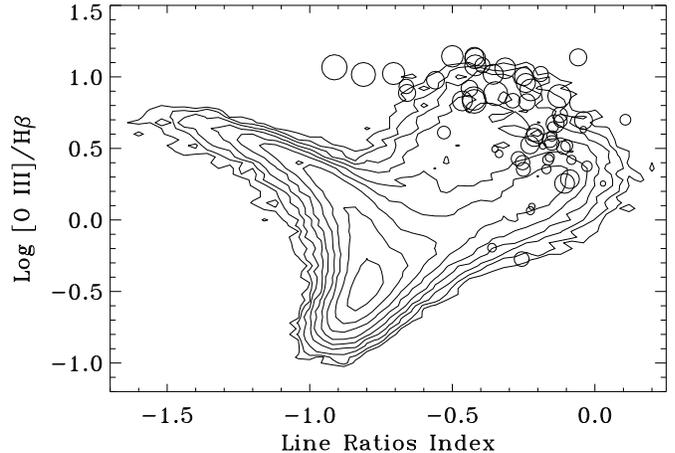,angle=90,width=1.0\linewidth}}
  \caption{Line Ratios Index (LRI, see text for its definition).  The contour
    lines indicate the distribution of the SDSS emission line galaxies
    \citepalias{kewley06b}. The circles indicate the 3CR sources (the symbol
    size is proportional to the [O~III] luminosity, see Fig. \ref{dd}).}
  \label{lri}
\end{figure}

In Fig. \ref{lri} we show the position of the 3CR sources in 
a new diagnostic plane where we compare log [O~III]/\Hb\ and LRI for the 59
narrow line objects for which we are able to measure both values.
The sources are distributed along the edges of the distribution of the SDSS
sources, similarly to the previous diagnostic diagrams. 
This diagram shows
two groups of AGN: one group with $-1\lesssim$LRI$\lesssim0$ and
log [O~III]/\Hb\ $\sim1$ and another one with $-0.3\lesssim$LRI$\lesssim0.1$
and $ 0.2\lesssim$ log [O~III]/\Hb$ \lesssim0.8$. 

\begin{figure}[htbp]
  \centerline{ \psfig{figure=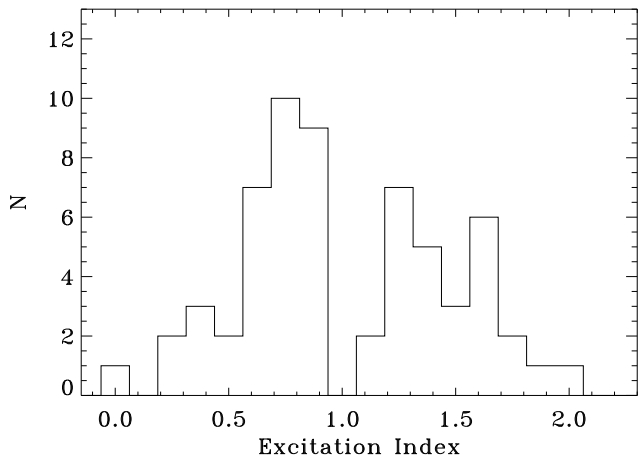,angle=90,width=1.0\linewidth}}
  \caption{ Histogram of the number of 3CR sources as function of the
    Excitation Index (see text)
representing the overall ratio of high and low excitation emission lines. 
Graphically, it is a coordinate axis parallel to the anti-bisectrix of Fig. \ref{lri}.}
  \label{exct}
\end{figure}

The presence of two populations of AGN is clearly seen in the distribution of
the values of the Excitation Index, shown in Fig. \ref{exct}, where two
separate distributions appear. This result is the analog of the two-horned
histograms derived by \citetalias{kewley06b} when slicing the density
distribution of SDSS sources. Using the KMM test for bimodality
\citep{ashman94}\footnote{available at
  {\it cas.umkc.edu/physics/ashman/blake.html}} we estimated that the
hypothesis of a single Gaussian distribution can be rejected at a confidence
level of 98\%. The significance level increases to 99.6\% when the two
galaxies with extremely low [O~III]/\Hb\ ratios are excluded from the
analysis. The estimated means of the two populations are E.I. = 0.63 and E.I.
= 1.40 for the low and high excitation sources respectively, with a standard
deviation of 0.25. We also note that the average rate of correct
classification of a galaxy within a given group is larger than 95\%.

\begin{figure}[hbtp]
  \centerline{ \psfig{figure=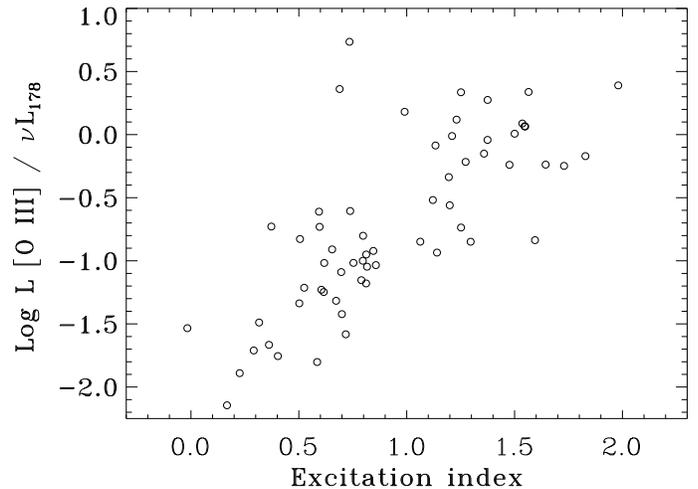,angle=90,width=1.15\linewidth}}
  \caption{Ratio of line and radio luminosities vs
    Excitation Index.}
  \label{o3-ei}
\end{figure}

An additional distinction between objects of high and low Excitation
Index is their radio emission.
We will discuss the connection between radio and optical properties in more
detail later on, but it is useful to anticipate this result. In Fig.
\ref{o3-ei} we compare the Excitation Index and 
$L_{\rm [OIII]}/\nu L_{178}$, i.e. the ratio between
their line and total radio (at 178 MHz) luminosity.
In this diagram, the two populations of galaxies 
at high and low values of E.I. 
are separated in two groups also from the point of view of the relative
level of line emission with respect to their radio luminosity.

The median values of log ($L_{\rm [OIII]}/\nu L_{178})$
are -0.1 for the galaxies with E.I. $\gtrsim 1$ and
-1.1 for lower excitation galaxies.

\begin{figure*}[t!]
  \centerline{\psfig{figure=13290f6a.epsi,angle=90,width=0.5\linewidth}
 \psfig{figure=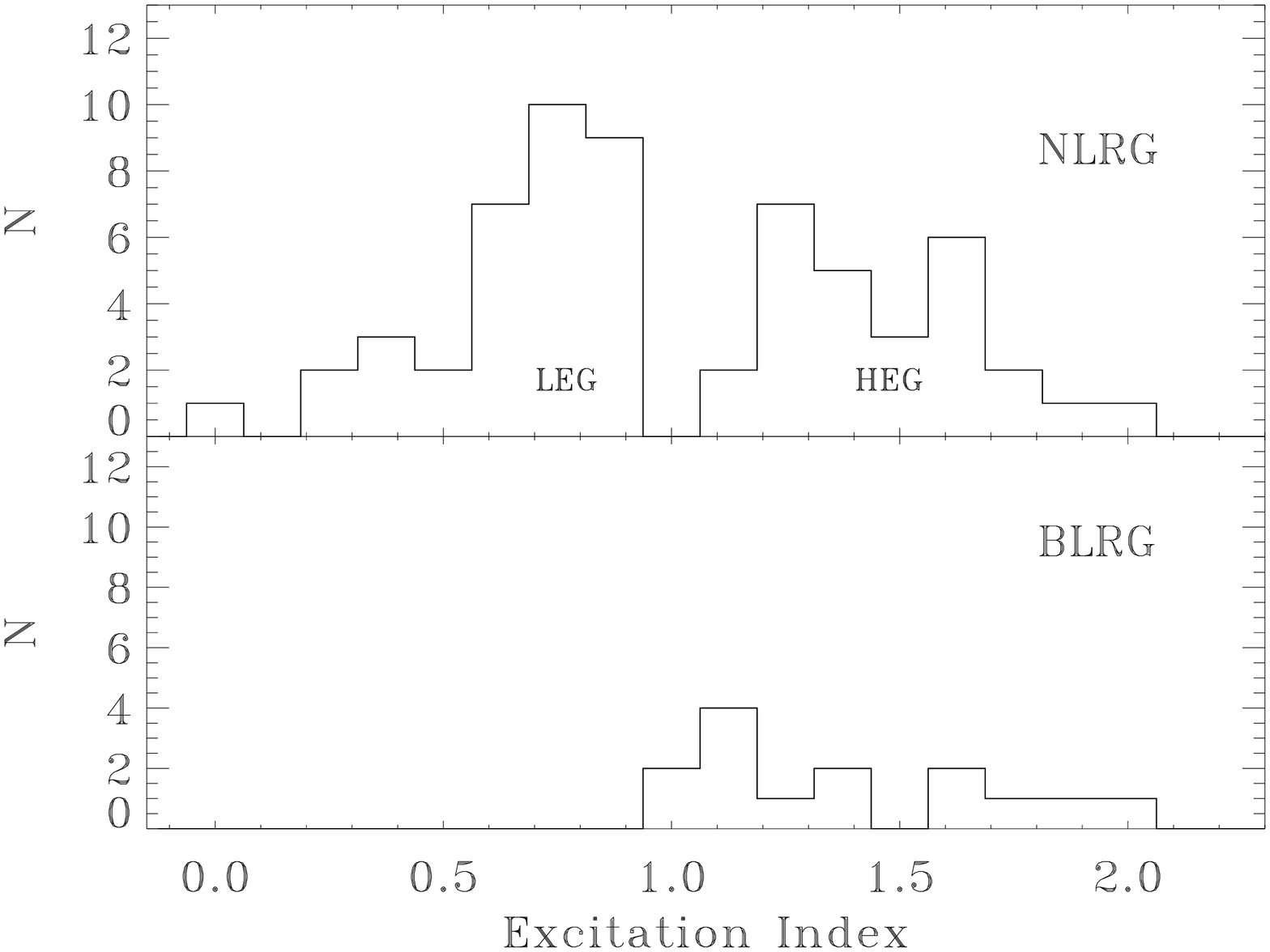,angle=90,width=0.5\linewidth}}
\caption{Left panel: LRI vs  log [O~III]/\Hb\ (as in Fig. \ref{lri}) 
including the broad line objects (crossed circles) and zooming onto
the 3CR populated region. 
HEG are represented by circles, LEG by squares. Right panel: comparison of the
distributions of Excitation Index for narrow and broad line objects.}
  \label{exct-blrg}
\end{figure*}

We conclude that RG can be separated into two sub-populations based on their
spectroscopic properties.  The threshold can be located at a value of the
Excitation Index of $\sim$ 0.95. This separation corresponds also to a
difference related to $L_{\rm [OIII]}/\nu L_{178}$ (the ratio between line and
radio luminosity) of a full order of magnitude. Following the terminology
introduced by \citet{laing94} we define these classes as Low and High
Excitation Galaxies, LEG and HEG respectively.

There are sources which remain unclassified using the
E.I. index. We will introduce alternative ways to classify these sources in
Sect. \ref{hegdd} and \ref{or}. 

\subsection{Spectroscopic properties of Broad Line Objects}
\label{blo}

We now include the broad line objects in our analysis\footnote{We exclude
  3C~234 from the sub-sample of Broad Lined Objects since for this source
  \citet{antonucci84} reports a high level of polarization ($\sim 14 \%$) for
  the broad emission line and continuum, at odds with the usually low level of
  polarization of Broad Line radio-galaxies. He interpreted this result as due
  to scattering of an otherwise hidden nucleus, at least in the optical band
  \citep{young98}. The observed broad line in this object represents only the
  small fraction of the intrinsic BLR emission that is reflected into our line
  of sight.}. In Fig. \ref{exct-blrg} (where we compare the Line Ratios Index
and the [O~III]/\Hb\ ratio) Broad Line Objects (BLO), marked with crossed
circles, are located in the region populated by HEG, although generally closer
to the separation line between HEG and LEG.  In fact, with respect to HEG, BLO
have a slightly lower average level of the [O~III]/\Hb\ ratio. As discussed
above this can be due to a general overestimate of the \Hb\ intensity in this
class of objects.  The distribution of Excitation Index of BLO is compared
with that of narrow line objects in Fig. \ref{exct-blrg}, right panel, and
again it shows a close overlap with high excitation galaxies. Although BLO
show a larger dispersion in E.I. (0.44) than HEG (0.24), the median E.I. of
the two classes differ by only $\sim$ 0.1.  Similarly, the $L_{\rm [OIII]}/\nu
L_{178}$ ratio of BLO is in the range 10$^{-1}$ - 10, the same covered by HEG.

Based on these results, we conclude that BLO can be considered, from the point
of view of their narrow lines intensity ratios and line over radio
luminosities, as members of the HEG class.

Including the BLO into the HEG class, the KMM test for bimodality still
rejects the hypothesis of a single Gaussian distribution at a confidence level
of 96.7 \% (still excluding the two galaxies with extremely low [O~III]/\Hb\
ratios).

\subsection{HEG and LEG in the diagnostic diagrams}
\label{hegdd}

\begin{figure*}[t!]
  \centerline{ \psfig{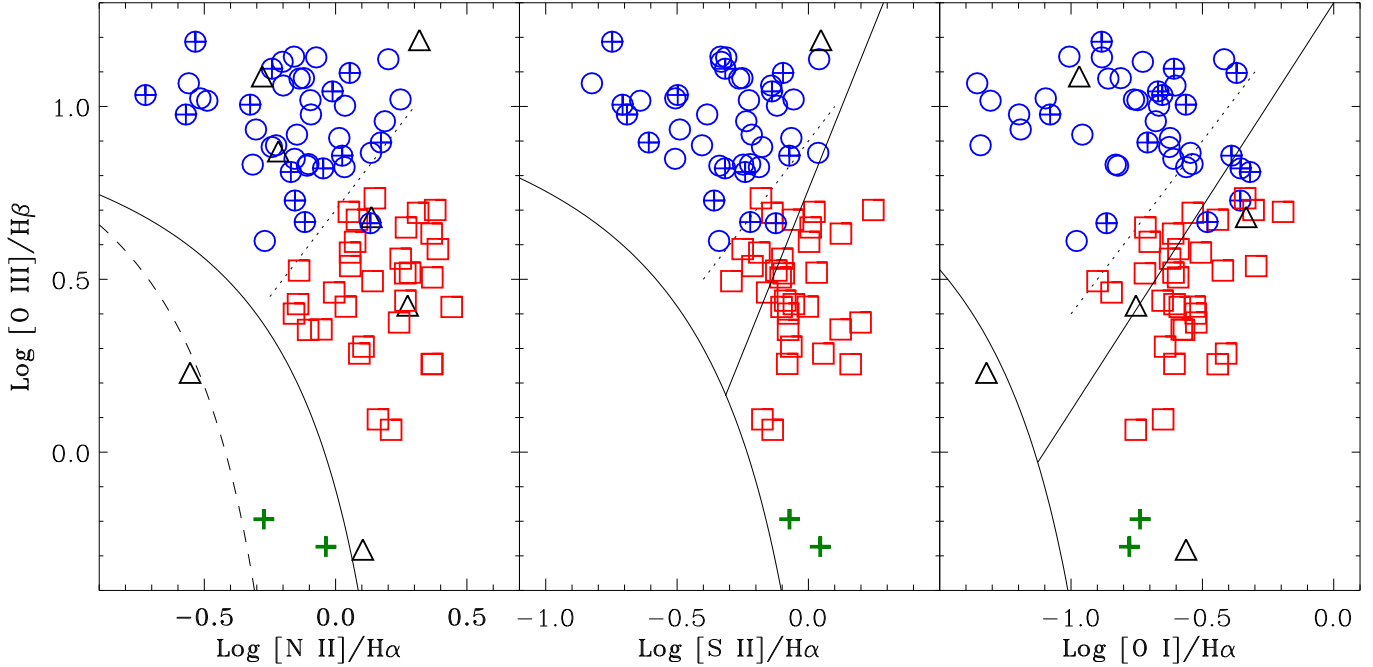}}
  \caption{ Diagnostic diagrams for 3CR sources after the classification into
    HEG (blue circles) and LEG (red squares) made using the Excitation Index.
    Crossed circles are broad line galaxies, green crosses are extremely low
    [O~III]/\Hb\ sources. Black triangles are sources for which the LRI
    cannot be estimated, as they lack the measurement of one or two 
diagnostic lines. The dotted lines mark the approximate boundaries
  between HEG and LEG in each diagram.}
  \label{diag-lri}
\end{figure*}

In Fig. \ref{diag-lri} we report the location of HEG (including 
Broad Line Objects) and LEG in the standard diagnostic spectroscopic
diagrams. As expected, HEG correspond generally to the objects with the
highest values of [O~III]/\Hb. They also have, on average, lower values of all
ratios of low excitation lines with respect to \Ha\ (i.e. they are located
to the top-left of LINERs). 
The two classes are completely separated in all diagrams
with the exception of at most two BLO that fall among the LEG. 
As already mentioned this is most likely due to an overestimate of the 
\Hb\ intensity and we note that a decrease of its luminosity by
20\% would be sufficient to move them among the Seyferts. 

In some cases it is possible to derive a rather robust classification also of
the 3CR sources for which not all the diagnostic lines could be measured.
These objects are represented by the black triangles. For example, two such
galaxies (namely 3C~381 and 3C~436) fall in the HEG regions, but they lack a
measurement of the [S II] or the [O I] line, due to the presence of a telluric
absorption band.  Based on their location in two of the diagrams a
classification as HEG appears secure.  In 3C~284, only upper limits can be
derived for the [S II] and [O I] fluxes; however, these limits confirm the
location of this object in the HEG region.  Similarly, we can give a LEG
classification for two sources, 3C~078 and 3C~357, whose spectra do not cover
the [S II] spectral region.

In addition, the sample comprises a single galaxy characterized by a star
forming spectrum (3C~198) and the three galaxies of extremely low excitation
(ELEG) discussed above, namely 3C~028, 3C~314.1, and 3C~348. The
spectroscopic classification of the sample is reported in Table \ref{crit}. In
Table \ref{speclas} we provide the classification for each object.

More generally, due to the relatively narrow stripe covered by the 3CR sources
in the diagnostic diagrams (and also by the relatively small number of
objects), it is difficult to define a two-dimensional representation for the
locus of the separation between LEG and HEG, such as those provided by
\citetalias{kewley06b}. As already mentioned, HEG and LEG are optimally
separated on the basis of a one-dimensional threshold in the excitation index
at E.I. $\sim$ 0.95.  However, when not all the required emission lines can be
measured, considering the individual diagrams we note that LEG are located 
approximately at

log [O~III]/\Hb\ - log  [N~II]/\Ha\ $\lesssim$ 0.7,

log [O~III]/\Hb\ - log  [S~II]/\Ha\  $\lesssim$ 0.9, and/or

log [O~III]/\Hb\ - log  [O~I]/\Ha\ $\lesssim$ 1.4.

\begin{table}
  \centering
  \caption{Spectral classification breakdown}
  \label{crit}
  \begin{tabular}{c| c c c c c| c}
    \hline \hline
     Method &HEG & HEG/BLO & LEG & ELEG &  SF & Total\\
    \hline
    E.I.    &27  &   14  &   32  &    2 &  --  &  75 \\
    D.D.    & 3  &   --  &    2  &    1 &   1  &   7 \\
    O.R.    &--  &    2  &    3  &   -- &  --  &   5 \\
    \hline                  
    Total   &30  &   16  &   37  &    3 &   1  &  87 \\
    \hline
  \end{tabular}
Classification method: E.I. - excitation index; D.D. - diagnostic diagrams; 
O.R. - emission line-radio correlation. 

\end{table}

\section{On  the line - radio relationship}
\label{or}

\subsection{Emission  lines vs radio luminosity}
\label{corr}

\begin{figure}[htbp]
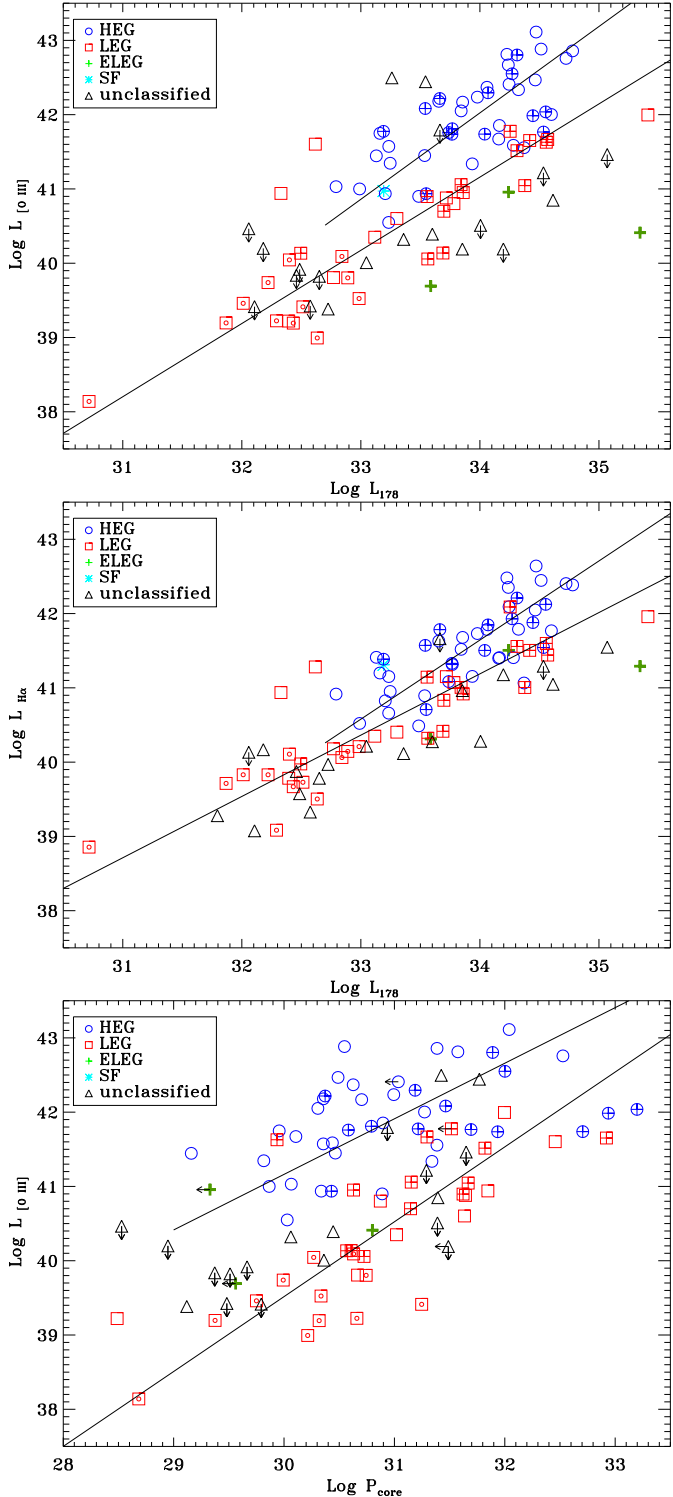

\centerline{\psfig{figure=13290f8a.epsi,angle=90,width=1.0\linewidth}}
\centerline{\psfig{figure=13290f8b.epsi,angle=90,width=1.0\linewidth}}
\centerline{\psfig{figure=13290f8c.epsi,angle=90,width=1.0\linewidth}}
\caption{\label{o3re} [O~III] and \Ha\ luminosity [erg s$^{-1}$] (top and
  middle panel, respectively) as a function of radio luminosity at 178 MHz [erg
  s$^{-1}$ Hz$^{-1}$]. Blue circles are HEG (crossed circles are BLO), red
  squares are LEG, green pluses are ELEG, the cyan asterisk is the star forming
  galaxy, while the black triangles are spectroscopically unclassified
  galaxies. The two solid lines represent the best linear fit obtained for the
  HEG and LEG sub-populations separately.  When possible, we further mark the
  LEG according to their FR type: crossed squares are FR~II/LEG and dotted
  squares are FR~I/LEG. Bottom panel: [O~III] luminosity as a function of
  radio core power.}
\end{figure}

In Fig. \ref{o3re}, top panel, we compare the [O~III] line luminosity with the
total radio luminosity at 178 MHz.  HEG are located in general at higher line
luminosity with respect to LEG of similar radio luminosity. This reflects
their higher value of the $L_{\rm [OIII]}/\nu L_{178}$ ratio already discussed
in Sect.\ref{hegleg}.

There are two notable exceptions to this trend, namely 3C~084 and 3C~371. They
are spectroscopically classified as LEG, but show an excess of a factor of
$\sim$ 50 in line emission with respect to the sources of this class of
similar radio power. Intriguingly, these are the two sources with the highest
core dominance of the sample, with ratios $P_{\rm core}$ / $L_{178}$ of 0.69
and 0.33 respectively. This implies that, by considering only their genuine
extended radio emission, they would be stronger outliers, reaching ratios of
line to radio emission even higher than observed in HEG.

While HEG are only found at radio luminosities larger than log $L_{178}$
[erg/s] $\gtrsim 32.8$, LEG cover the whole range of radio power covered by the
subsample of 3CR sources with z$<$0.3, almost five orders of magnitude from
log $L_{178}\sim30.7$ to log $L_{178}\sim35.4$.

There is a clear trend for increasing line luminosity with radio
power, as already found and discussed by several authors and as reported in
the introduction.  However, we are now in the position of considering
separately the sub-populations of HEG and LEG.  More quantitatively, we find
that HEG obey a linear correlation in the form:

log $L_{\rm[O~III]}$ = 1.15 $~$log $L_{178}$ + 2.96. 

The error in the slope is 0.11, while the rms around the correlation is 0.43
dex.  We tested the possible influence of redshift in driving this correlation
(both quantities depend on $z^2$) estimating the partial rank coefficient,
finding $r_{{\rm [O III]} - L_{178},z}=0.41$. For 46 data-points, the
probability that this results from randomly distributed data is $P = 0.005$.
For LEG we find:

log $L_{\rm[O~III]}$ = 0.99 $~$log $L_{178}$ + 7.65. 

The error in the slope is 0.09 and the rms around the correlation is 0.50 dex.
The partial rank coefficient is $r_{{\rm [O III]} - L_{178},z}=0.51$,
corresponding to a probability of a random distribution of $P = 0.002$.
 
The relations derived for HEG and LEG differ by a factor of $\sim$ 10 in the
common range of radio power.  The slopes of the line-radio luminosity
correlations of HEG and LEG are instead consistent within the errors.

Considering instead the \Ha\ line, we find rather similar results, as shown in
Fig. \ref{o3re}, middle panel. The form of the correlations are:

log $L_{\rm {H\alpha}}$ = 1.06 $~$log $L_{178}$ + $\,\,\,$5.44 (for HEG) and

log $L_{\rm {H\alpha}}$ = 0.83 $~$log $L_{178}$ + 13.13 (for LEG).

The main difference is the smaller offset between
the two populations, reduced to a factor of $\sim$ 3. Also the spread among
the two classes is reduced to 0.22 and 0.44 dex for HEG and LEG, respectively.

Finally, correlations are found also between radio core power and line
luminosity (Fig. \ref{o3re}, bottom panel) with

log $L_{\rm[O III]}$ = 0.75 $~$$P_{\rm core}$ + $\,\,\,$18.71 (for HEG) and

log $L_{\rm[O III]}$ = 1.01 $~$$P_{\rm core}$ + 9.28 (for LEG).

The separation in terms of ratio between line and radio luminosity provides us
with a further element to associate spectroscopically unclassified sources with
the HEG and LEG sub-populations. For example, the narrow Balmer lines of two
BLO, 3C~111 and 3C~445, could not be measured, due to the prominence of the
broad component. Nonetheless, they are located at log $L_{178}\sim$33.5 and
log $L_{\rm [O III]}\sim$42.5, so the line luminosity factor is $\sim$ 100 times
higher than LEG of similar radio luminosity, suggesting an identification of
these two sources as HEG.  Adopting this definition, all BLO belong to the HEG
population.

Conversely, the nature of the relatively large population of unclassified
sources remains ambiguous. In general, they show values (or upper limits) of
the $L_{\rm [OIII]}/\nu L_{178}$ ratios significantly smaller than predicted
by the line-radio correlation defined by HEG. However, a LEG classification is
not granted since we cannot exclude that they are part of the ELEG
population, as they lack a measurement of the [O~III]/\Hb\ ratio.  While the
\Hb\ flux is never available, in a few cases the [O~III] line can be measured
leading to a ratio log [O~III]/\Hb $>$ 0.1, inconsistent with a ELEG
classification.  This is the case of 3C~173.1, 3C~326, and 3C~430
which may then be considered as LEG.

\subsection{Spectroscopic classes vs radio morphology}
\label{egfr}
We now examine the relationship between the spectroscopic classes and the
radio morphological FR type. In Table \ref{speclas} we report our own
classification of the 3CR radio-sources based on the analysis of the best
radio maps available in the literature. We adopted rather strict criteria for
the inclusion of a given object into one of the FR morphological type,
preferring a larger number of undefined sources with respect to a less secure
identification. In particular, we include a radio source into the FR~II group
only when clear hot spots are visible in the radio maps, and considered as
FR~I only sources with a well defined twin-jets structure.

All sources of the HEG spectroscopic class belong to the FR~II type,
with only two exceptions. The first is 3C~433 that shows a peculiar radio
morphology. The South radio lobe has the presence of a well defined
hot-spot, while the Northern radio jets bends dramatically toward the West,
forming a rather diffuse lobe \citep{black92}.  The second is 3C~93.1, a
Compact Steep Spectrum source \citep{akujor91} too compact ($\sim 0\farcs2$ in
size) to show any structure. As already noted, 
all HEG are found in radio-sources with log $L_{178}\gtrsim32.8$.

The situation for LEG is more complex. Of the 37 sources belonging to this
spectroscopic class 16 are FR~II, 12 are FR~I, while the radio-morphology of
the remaining 9 is not sufficiently well defined to include them in any FR
group, such as, for example, 3C~317 (a core/halo source), 3C~371 (with a
core/jet morphology), and the X-shaped 3C~315.

LEG with FR~II morphology are associated with radio-sources with relatively
large radio power, from log $L_{178}\sim33.5$ to log $L_{178}\gtrsim34.5$,
with just one exception (3C~088) of lower luminosity, while the FR~I/LEG all
have log $L_{178}\lesssim33$. We also stress that the excitation level of
  FR~II/LEG (with a mean value of E.I. = 0.66 and a spread of 0.15) does not
  differ from the rest of the LEG population (for which the mean is E.I. =
  0.63). This indicates that radio
  morphology does not affect the excitation levels.

For the objects not classified spectroscopically, we have 10 FRI, 4 FRII and 3
of uncertain FR type.

Conversely, by looking at the spectral classification of the different FR
types, while there are FR~II/HEG as well as FR~II/LEG, all FR~I for
which we were able to derive a spectral type are LEG.

\begin{table*}
  \begin{center}
    \caption{Multiwavelength data and spectroscopic classification}
    \label{speclas}
    \begin{tabular}{l| c| c c| c c| c| c c c}
      \hline \hline
Name & redshift & \multicolumn{2}{|c|}{ Emission lines}& \multicolumn{2}{|c|}{
  Radio emission}  & Host magnitude & \multicolumn{3}{|c}{Classification}  \\
\hline
     & &  H$\alpha$  & [O~III] & L$_{178}$ &  P$_{core}$ & M$_{H}$ &  FR & spec & Method\\ 
\hline 
3C~015   &  0.073    &   40.40    &  40.60  &   33.30 &  31.64 &  -25.29    &   & LEG  &E.I.\\
3C~017   &  0.2198   &   41.88    &  41.99  &   34.44 &  32.94 &  -24.81$^*$& 2 & BLO  &E.I.\\
3C~018	 &  0.188    &   41.93    &  42.55  &   34.27 &  32.00 &    --      & 2 & BLO  &E.I.\\
3C~020   &  0.174    &\multicolumn{2}{|c|}{no obs.}&   34.55 &  30.44 &  -24.64$^*$& 2  &\multicolumn{2}{c}{no obs.}\\
3C~028	 &  0.1952   &   41.51    &  40.96  &   34.24 &  29.33 &    --      &   & ELEG  &E.I.\\
3C~029	 &  0.0448   &   40.06    &  40.09  &   32.84 &  30.63 &  -25.44    & 1 & LEG  &E.I.\\
3C~031   &  0.0167   &   39.83    &  39.46  &   32.01 &  29.75 &  -25.51    & 1 & LEG  &E.I.\\
3C~033	 &  0.0596   &   41.63    &  42.18  &   33.65 &  30.36 &  -24.75    & 2 & HEG  &E.I.\\
3C~033.1 &  0.1809   &   41.85    &  42.30  &   34.07 &  31.19 &  -24.47$^*$& 2 & BLO  &E.I.\\
3C~035	 &  0.0670   &   40.22    &  40.01  &   33.05 &  30.36 &  -25.17    & 2 & --   &\\
3C~040	 &  0.0185   &   39.08    &  39.22  &   32.29 &  30.66 &    --      & 1 & LEG  &E.I.\\
3C~052	 &  0.2854   & $<$40.64   &  --     &   34.53 &  31.29 &  -26.74$^*$& 2 & --   &\\
3C~061.1 &  0.184    &   42.05    &  42.47  &   34.47 &  30.49 &  -23.50$^*$& 2 & HEG  &E.I.\\
3C~063   &  0.175    &\multicolumn{2}{|c|}{no obs.}&   34.21 &  31.12 &    --      &   & \multicolumn{2}{c}{no obs.}\\
3C~066B	 &  0.0215   &   40.11    &  40.05  &   32.40 &  30.27 &  -26.25$^*$& 1 & LEG  &E.I.\\
3C~075N	 &  0.0232   &   39.58    & $<$39.92&   32.49 &  29.67 &  -24.51$^*$& 1 & --   &\\
3C~076.1 &  0.0324   &   39.89    & $<$39.85&   32.46 &  29.37 &  -24.08$^*$& 1 & --   &\\
3C~078	 &  0.0286   &   39.73    &  39.41  &   32.51 &  31.25 &  -26.16    & 1 & LEG  &D.D.\\
3C~079	 &  0.2559   &   42.39    &  42.86  &   34.78 &  31.39 &  -25.27    & 2 & HEG  &E.I.\\
3C~083.1 &  0.0255   &   39.40    & $<$39.50&   32.57 &  29.48 &  -26.70$^*$& 1 & --   &\\
3C~084	 &  0.0176   &   41.28    &  41.60  &   32.62 &  32.46 &  -25.99    &   & LEG  &E.I.\\
3C~088	 &  0.0302   &   39.98    &  40.14  &   32.49 &  30.57 &  -24.81    & 2 & LEG  &E.I.\\
3C~089   &  0.1386   &   40.28    & $<$40.51&   34.01 &  31.39 &  -26.22    & 1 & --   &\\
3C~093.1 &  0.2430   &   42.35    &  42.67  &   34.24 &   --   &    --      &  & HEG  &E.I.\\
3C~098   &  0.0304   &   40.52    &  41.00  &   32.99 &  29.87 &  -24.38    & 2 & HEG  &E.I.\\
3C~105	 &  0.089    &   40.89    &  41.45  &   33.54 &  30.46 &  -24.31    & 2 & HEG  &E.I.\\
3C~111	 &  0.0485   &    --      &  42.44  &   33.54 &  31.77 &  -25.07    & 2 & BLO  &O.R.\\
3C~123	 &  0.2177   &   41.96    &  42.00  &   35.41 &  32.00 &  -26.58$^*$&   & LEG  &E.I.\\
3C~129	 &  0.0208   &   39.81    & $<$39.85&   32.65 &  29.51 &  -25.11    & 1 & --   &\\
3C~129.1 &  0.0222   &  $<$39.83  &  --     &   32.06 &  28.53 &  -25.61    & 1 & --   &\\
3C~130	 &  0.1090   &  $<$40.17  &  --     &   33.66 &  30.94 &  -28.45    & 1 & --   &\\
3C~132   &  0.214    &\multicolumn{2}{|c|}{no obs.}&   34.25 &  31.58 &  -26.00    & 2 & \multicolumn{2}{c}{no obs.}\\
3C~133	 &  0.2775   &   42.41    &  42.76  &   34.72 &  32.53 &  -25.36$^*$& 2 & HEG  &E.I.\\
3C~135	 &  0.1253   &   41.52    &  42.05  &   33.84 &  30.31 &  -24.47    & 2 & HEG  &E.I.\\
3C~136.1 &  0.064    &   41.41    &  41.44  &   33.13 &  29.16 &  -25.17    & 2 & HEG  &E.I.\\
3C~153   &  0.2769   &   41.60    &  41.63  &   34.56 &  29.94 &  -25.60$^*$& 2 & LEG  &E.I.\\
3C~165	 &  0.2957   &   41.44    &  41.67  &   34.57 &  31.30 &  -25.80$^*$& 2 & LEG  &E.I.\\
3C~166	 &  0.2449   &   41.51    &  41.66  &   34.42 &  32.92 &  -25.32$^*$& 2 & LEG  &E.I.\\
3C~171	 &  0.2384   &   42.45    &  42.89  &   34.51 &  30.55 &  -24.73$^*$& 2 & HEG  &E.I.\\
3C~173.1 &  0.2921   &   41.05    &  40.85  &   34.61 &  31.39 &  -26.48    & 2 & LEG  &O.R.\\
3C~180   &  0.22     &   41.79    &  42.34  &   34.32 &   --   &  -24.94    & 2 & HEG  &E.I.\\
3C~184.1 &  0.1182   &   41.79    &  42.23  &   33.66 &  30.37 &  -24.22$^*$& 2 & BLO  &E.I.\\
3C~192	 &  0.0598   &   40.95    &  41.34  &   33.25 &  29.82 &  -24.68    & 2 & HEG  &E.I.\\
3C~196.1 &  0.198    &   41.56    &  41.52  &   34.31 &  31.82 &  -25.47    & 2 & LEG  &E.I.\\
3C~197.1 &  0.1301   &   40.69    &  40.92  &   33.55 &  30.43 &  -24.94    & 2 & BLO  &E.I.\\
3C~198	 &  0.0815   &   41.31    &  40.97  &   33.19 &   --   &  -23.62$^*$&   & SF   &D.D.\\
3C~213.1 &  0.194    &   41.01    &  41.06  &   33.84 &  31.15 &  -25.02$^*$& 2 & LEG  &E.I.\\
3C~219	 &  0.1744   &   41.55    &  41.77  &   34.53 &  31.69 &  -25.70    & 2 & BLO  &E.I.\\
3C~223	 &  0.1368   &   41.68    &  42.17  &   33.85 &  30.70 &  -24.74    & 2 & HEG  &E.I.\\
3C~223.1 &  0.107    &   41.16    &  41.58  &   33.23 &  30.36 &  -24.95    & 2 & HEG  &E.I.\\
3C~227	 &  0.0861   &   41.08    &  41.75  &   33.74 &  30.58 &  -24.90    & 2 & BLO  &E.I.\\
3C~234   &  0.1848   &   42.64    &  43.11  &   34.47 &  32.04 &  -26.09    & 2 & HEG  &E.I.\\
3C~236	 &  0.1005   &   41.13    &  40.89  &   33.56 &  31.62 &  -25.34    & 2 & LEG  &E.I.\\
3C~258   &  0.165    &   40.96    &  40.19  &   33.85 &   --   &    --      &   & --   &\\
3C~264   &  0.0217   &   39.68    &  39.20  &   32.43 &  30.32 &  -25.09    & 1 & LEG  &E.I.\\
3C~272.1 &  0.0035   &   38.92    &  38.20  &   30.72 &  28.68 &  -24.43    & 1 & LEG  &E.I.\\
3C~273   &  0.1583   &    --      &  --     &   34.62 &  33.65 &    --      &   & BLO  &\\
3C~274	 &  0.0044   &   39.50    &  38.99  &   32.63 &  30.21 &  -25.28    & 1 & LEG  &E.I.\\
\hline                                                                     
  \multicolumn{9}{c}{{Continued on Next Page}} \\                        
    \end{tabular}                                
  \end{center}                                   
\end{table*}                                     
                                                 
\addtocounter{table}{-1}                         
\begin{table*}
  \begin{center}
    \caption{Continued}
    \begin{tabular}{l| c| c c| c c| c| c c c}
      \hline \hline
Name & redshift & \multicolumn{2}{|c|}{ Emission lines}& \multicolumn{2}{|c|}{
  Radio emission}  & Host magnitude & \multicolumn{3}{|c}{Classification}  \\
\hline
     & &  H$\alpha$  & [O~III] & L$_{178}$ &  P$_{core}$ & M$_{H}$ &  FR & Class & Method\\ 
 \hline 
3C~277.3 &  0.0857   &   40.83   &  40.94  & 33.21 &   30.34 &  -24.87    & 2 & HEG  &E.I.\\
3C~284   &  0.2394   &   41.41   &  41.59  & 34.28 &   30.44 &  -25.57    & 2 & HEG  &D.D.\\
3C~285	 &  0.0794   &   40.66   &  40.55  & 33.23 &   30.03 &  -24.53    & 2 & HEG  &E.I.\\
3C~287.1 &  0.2159   &   41.50   &  41.73  & 34.04 &   32.71 &  -25.72    & 2 & BLO  &E.I.\\
3C~288   &  0.246    &\multicolumn{2}{|c|}{no obs.}& 34.53 &   31.73 &  -26.10$^*$& 2 & \multicolumn{2}{c}{no obs.}\\
3C~293   &  0.0450   &   40.18   &  39.80  & 32.77 &   30.67 &  -25.33    &   & LEG  &E.I.\\
3C~296   &  0.0240   &   39.87   &  39.78  & 32.22 &   29.99 &  -26.04    & 1 & LEG  &E.I.\\
3C~300   &  0.27     &   41.78   &  42.01  & 34.60 &   31.27 &  -24.92    & 2 & HEG  &E.I.\\
3C~303	 &  0.141    &   41.33   &  41.74  & 33.77 &   31.94 &  -25.35    & 2 & BLO  &E.I.\\
3C~303.1 &  0.267    &   42.10   &  42.42  & 34.25 &   31.04 &    --      & 2 & HEG  &E.I.\\
3C~305	 &  0.0416   &   40.92   &  41.03  & 32.79 &   30.07 &  -25.26    & 2 & HEG  &E.I.\\
3C~310	 &  0.0535   &   40.32   &  40.05  & 33.56 &   30.72 &  -25.02$^*$& 2 & LEG  &E.I.\\
3C~314.1 &  0.1197   &   40.31   &  39.69  & 33.59 &   29.56 &    --      &   & ELEG  &E.I.\\
3C~315   &  0.1083   &   41.15   &  40.87  & 33.72 &   31.64 &  -24.74$^*$&   & LEG  &E.I.\\
3C~317	 &  0.0345   &   40.35   &  40.35  & 33.12 &   31.02 &  -26.04    &   & LEG  &E.I.\\
3C~318.1 &  0.0453   &   39.95   &  39.36  & 32.72 &   29.12 &  -25.70    &   & --   &\\
3C~319   &  0.192    &   41.16   & $<$40.16& 34.20 &   31.49 &  -24.41$^*$& 2 & --   &\\
3C~321   &  0.096    &   40.50   &  40.91  & 33.49 &   30.89 &  -25.52    & 2 & HEG  &E.I.\\
3C~323.1 &  0.264    &   42.21   &  42.80  & 34.31 &   31.89 &  -26.74    & 2 & BLO  &E.I.\\
3C~326	 &  0.0895   &   40.28   & 40.40   & 33.60 &   30.45 &  -24.33    & 2 & LEG  &O.R.\\
3C~327   &  0.1041   &   41.73   &  42.24  & 33.98 &   30.99 &    --      & 2 & HEG  &E.I.\\
3C~332   &  0.1517   &   41.31   &  41.81  & 33.77 &   30.79 &  -25.38    & 2 & BLO  &E.I.\\ 
3C~338   &  0.0303   &   40.25   &  39.57  & 32.99 &   30.34 &  -26.21$^*$& 1 & LEG  &E.I.\\
3C~346   &  0.161    &\multicolumn{2}{|c|}{no obs.}& 33.88 &   32.18 &  -25.84    &  2 &\multicolumn{2}{c}{no obs.}\\
3C~348   &  0.154    &   41.29   &  40.40  & 35.35 &   30.80 &    --      &   & ELEG  &D.D.\\
3C~349   &  0.205    &\multicolumn{2}{|c|}{no obs.}& 34.20 &   31.35 &  -24.82$^*$& 2 &\multicolumn{2}{c}{no obs.}\\
3C~353	 &  0.0304   &   40.42   &  40.14  & 33.69 &   30.61 &  -24.77    & 2 & LEG  &E.I.\\
3C~357   &  0.1662   &   40.92   &  40.95  & 33.86 &   30.63 &  -25.83$^*$& 2 & LEG  &D.D.\\
3C~371   &  0.0500   &   40.94   &  40.94  & 32.33 &   31.85 &  -25.36    &   & LEG  &E.I.\\
3C~379.1 &  0.256    &   41.41   &  41.86  & 34.16 &   30.90 &  -25.69    & 2 & HEG  &E.I.\\
3C~381   &  0.1605   &   41.79   &  42.37  & 34.06 &   30.63 &  -24.81    & 2 & HEG  &D.D.\\
3C~382   &  0.0578   &   41.39   &  41.78  & 33.19 &   31.22 &  -26.03    & 2 & BLO  &E.I.\\
3C~386	 &  0.0170   &   40.17   & $<$40.20& 32.18 &   28.95 &  -24.57    &   & --   &\\
3C~388   &  0.091    &   40.83   &  40.70  & 33.70 &   31.15 &  -26.20    & 2 & LEG  &E.I.\\
3C~390.3 &  0.0561   &   41.57   &  42.08  & 33.54 &   31.46 &  -24.84    & 2 & BLO  &E.I.\\
3C~401   &  0.2010   &   41.01   &  41.05  & 34.38 &   31.67 &  -25.03$^*$& 2 & LEG  &E.I.\\
3C~402	 &  0.0239   &   39.08   & $<$39.42& 32.11 &   29.79 &  -24.77$^*$& 1 & --   &\\
3C~403	 &  0.0590   &   41.20   &  41.75  & 33.16 &   29.96 &  -25.27    & 2 & HEG  &E.I.\\
3C~403.1 &  0.0554   &\multicolumn{2}{|c|}{no obs.}& 32.98 &    --   &  -24.36    &   &\multicolumn{2}{c}{no obs.}\\
3C~410   &  0.2485   &\multicolumn{2}{|c|}{no obs.}& 34.80 &   33.43 &    --      & 2 &\multicolumn{2}{c}{no obs.}\\
3C~424   &  0.127    &   41.07   &  40.80  & 33.78 &   30.87 &  -23.96$^*$&   & LEG  &E.I.\\
3C~430   &  0.0541   &   40.12   &  40.33  & 33.36 &   30.06 &  -25.28    & 2 & LEG  &O.R.\\
3C~433	 &  0.1016   &   41.40   &  41.67  & 34.16 &   30.11 &  -25.79$^*$&   & HEG  &E.I.\\
3C~436   &  0.2145   &   41.07   &  41.56  & 34.37 &   31.39 &  -25.50    & 2 & HEG  &D.D.\\
3C~438   &  0.290    &   41.55   & $<$41.46& 35.07 &   31.65 &  -26.57    & 1 & --   &\\
3C~442	 &  0.0263   &   39.78   &  39.21  & 32.39 &   28.49 &    --      &   & LEG  &E.I.\\
3C~445   &  0.0562   &    --     &  42.50  & 33.26 &   31.42 &    --      & 2 & BLO  &O.R.\\
3C~449	 &  0.0171   &   39.71   &  39.19  & 31.87 &   29.38 &  -24.80$^*$& 1 & LEG  &E.I.\\
3C~452	 &  0.0811   &   41.16   &  41.34  & 33.94 &   31.34 &  -24.92    & 2 & HEG  &E.I.\\
3C~456   &  0.2330   &   42.48   &  42.81  & 34.23 &   31.57 &    --      & 2 & HEG  &E.I.\\
3C~458   &  0.2890   &\multicolumn{2}{|c|}{no obs.}& 34.58 &   30.88 &    --      & 2 &\multicolumn{2}{c}{no obs.}\\
3C~459	 &  0.2199   &   42.17   &  42.03  & 34.55 &   33.20 &  -25.34$^*$& 2 & BLO  &E.I.\\
3C~460   &  0.268    &   42.09   &  41.78  & 34.25 &   31.52 &    --      & 2 & LEG  &E.I.\\
3C~465   &  0.0303   &   40.15   &  39.81  & 32.89 &   30.74 &  -26.44$^*$& 1 & LEG  &E.I.\\
\hline                                                          
    \end{tabular}                                               
  \end{center}                                                  
  Column description: (1) 3CR name; (2) redshift from \citet{spinrad85}; (3) and (4) logarithm
  of H$\alpha$ and [O~III]$\lambda$5007 luminosities [erg s$^{-1}$] from \citet{buttiglione09}; (5) radio luminosity at
  178 MHz [erg s$^{-1}$ Hz$^{-1}$] from \citet{spinrad85}; (6) radio core power at 5 GHz [erg
  s$^{-1}$ Hz$^{-1}$] from \citet{baldi09}; (7) host
  H magnitude from 2MASS \citep{skrutskie06} 
(or from HST \citep{donzelli07} for the objects marked with a $^*$);
  (8): morphological FR type; (9) spectroscopic classification into High
  Excitation Galaxy (HEG); 
  Low Excitation Galaxy (LEG); Broad Line Object (BLO); Extremely Low
  Excitation Galaxy (ELEG); (SF) starforming galaxy; (--) unclassified.  Column (10)
  classification method: E.I. - excitation index; D.D. - diagnostic diagrams; O.R.
  - emission line radio correlation. 
\end{table*}

\subsection{Broad emission lines and spectroscopic classes}
\label{blrsep}

\begin{figure}[htbp]
  \centerline{ 
    \psfig{figure=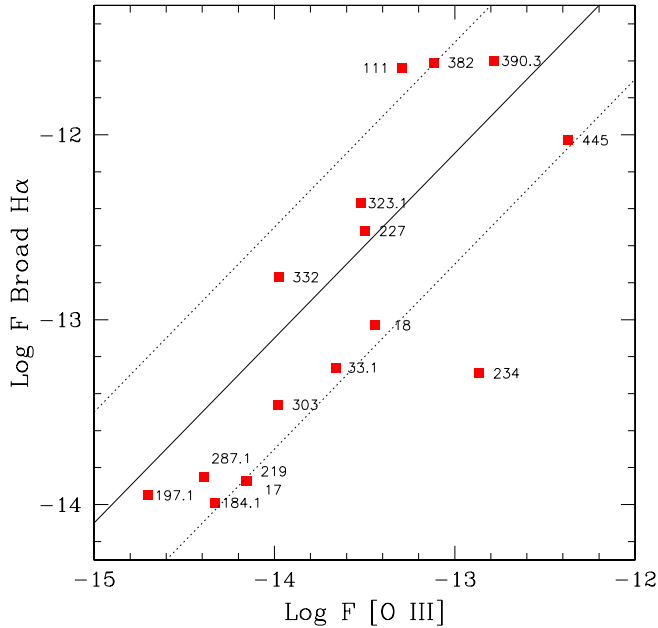,width=1.00\linewidth}}
  \caption{\label{fo3blr} Comparison of the broad H$\alpha$ and [O III] line
    fluxes for 17 RG showing a broad emission line component (we
    excluded 3C~273 since it lacks the [O III] flux). The solid line
    represents the locus of galaxies with $F_{{\rm H\alpha(broad})}$ = 8
    $F_{\rm[O III]}$. The dotted lines show the effect of varying this ratio
    by a factor 4. The outlier is 3C~234, whose broad lines are due to
    scattered light.}
\end{figure}

Broad \Ha\ lines are found in a sub-sample of 18 3CR sources.  With the
possible exception of 3C~273, where no narrow lines can be measured, all
remaining sources are part of the HEG sub-population. The broad emission line
fluxes are linked by a linear relation to the [O III] flux over 2.5 orders of
magnitude, with an average ratio $F_{{\rm H\alpha(broad})} \sim 8 F_{\rm[O
  III]}$, and a scatter of a factor of $\sim$ 4 (see Fig.
\ref{fo3blr}).\footnote{The only outlier from this trend is 3C~234, already
  discussed previously, and in which the observed broad lines are due to
  scattered light.}  This is expected since both lines are essentially
proportional to the flux of ionizing photons. Differences in their relative
intensities, from object to object, are most likely due to changes in the
relative covering factor of the Narrow and Broad Line Regions.

Under the assumption that the flux of the \Ha\ broad component follows the
relation with [O~III] of Fig. \ref{fo3blr} and with similar
properties in terms of width and shape of the profile, one would have expected
to detect a BLR in the objects with $F_{\rm{[O~III] }} \gtrsim 2 \times
10^{-15}$ erg s$^{-1}$ cm$^{-2}$, 
the lowest [O III] flux among BLO, i.e. in $\sim$ 75\% 
of the sample. This indicates that for these galaxies the $F_{{\rm
    H\alpha(broad})} / F_{\rm[O III]}$ ratio must be lower than in BLO. For
HEG lacking of a BLR this can be accounted for by selective absorption, i.e. to
circumnuclear obscuring material, as predicted by the unified models
\citep[e.g.][]{urry95}. However, a BLR is never observed in LEG. This is
difficult to account in a simple geometric scheme and it implies that BLR in
LEG are intrinsically fainter, relative to the narrow line emission, than in
HEG. 

\subsection{Comparison with previous studies}
\label{previous} 

We based our initial classification on the ratios of diagnostic emission
lines. Previous studies used instead a combination of line ratios and
equivalent widths (see the Introduction) that can be more affected (in
particular when only upper limits can be derived) by the quality of the data
and by the contrast with the continuum emission. Similarly, we only used
ratios of lines with small wavelength separation, not affected by the possible
effects of internal reddening, that can be particularly severe when
considering e.g. the [O II]$\lambda$3727 line. Thus our procedure is expected
to produce a rather robust method of spectral identification.

Nonetheless, our classifications are overall in good agreement with those
found in the literature on a object by object basis.  For example, comparing
our results with those of \citet{willott99}\footnote{available at\\
{\it http://www.science.uottawa.ca/$\sim$cwillott/3crr/3crr.html}}
for the 3CRR sources we found 52 objects in common. Leaving aside 2 objects of
the newly introduced class of ELEG, and 3 objects that we consider as
unclassified (2 reported as LEG, namely 3C~035 and 3C~319, 1 as HEG, 3C~438)
the identification in the various classes coincides with only 3 exceptions for
the remaining 47 radio-galaxies. These are: 3C~388, a LEG from our analysis
(with an excitation index of E.I.=0.62) against the previous HEG
identification, and two galaxies, 3C~079 and 3C~223, where we do not see a
broad line component\footnote{We estimated a limit to their broad \Ha\ flux of
  1.6$\times 10^{-15}$ and 1.8$\times 10^{-16} \ergscm$ for 3C~079 and 3C~223
  respectively. Given their [O~III] fluxes, these limits correspond to a
  factor of 200-300 lower than the mean value $F_{{\rm H\alpha(broad})}$ = 8
  $F_{\rm[O III]}$ reported in Sect.  \ref{blrsep} (see also Fig.
  \ref{fo3blr}).}, contrasting with their suggested membership in the class of
Weak Quasars.

\subsection{The radio - host galaxy luminosity plane for 
FR I, FR II, LEG,  and HEG}
\label{frsep}

\begin{figure*}[htbp]
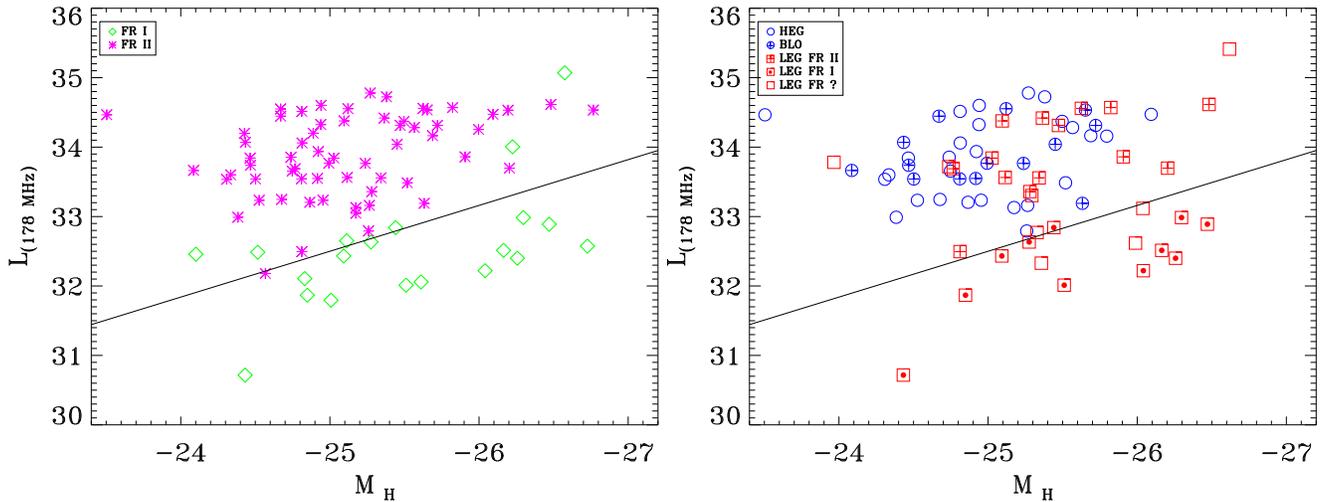

  \centerline{ 
    \psfig{figure=13290f10.epsi,width=0.48\linewidth,angle=90}
    \psfig{figure=13290f11.epsi,width=0.48\linewidth,angle=90}
}
\caption{\label{ledlow} Comparison of the radio luminosity at 178 MHz versus
  the magnitude in H band of the host galaxy for the various sub-populations
  of 3CR sources. Left panel: 3CR sources divided into FR~I (green diamonds)
  and FR~II (pink stars); right panel: comparison between HEG (blue circles)
  and LEG (FR~I = crossed red squares, LEG FR~II = filled red squares,
  LEG of uncertain FR type = empty red squares).
  The solid line marks the separation between FR~I (below the line) and FR~II
  (above the line) according to \citet{ledlow96}.}
\end{figure*}

\citet{ledlow96} compared the optical R band magnitude of the host galaxies
with the total radio emission at 1.4 GHz.  They found that sources locate in
different areas of the plot depending on their radio morphology: as already
known from the pioneering study of \citet{fanaroff74} FR~II sources have
higher radio powers than FR~I sources and they separate at a luminosity of
$\sim 2\times10^{25}$ W Hz$^{-1}$ at 178 MHz.  The novel result of
\citet{ledlow96} consists in the fact that the FR~I/II division shows a
dependence on $M_{\rm host}$.  FR~I sources hosted by the more luminous
galaxies can have radio powers higher than the average FR~I/FR~II separation.
The separation between FR~I and FR~II is rather sharp over the whole range of
radio power.

In Fig. \ref{ledlow} (left panel) we plotted the 3CR sources in the plane
radio luminosity (at 178 MHz) versus the magnitude in H band of the host
galaxy (reported in Table \ref{speclas}). We selected the H band since it
provides the most complete coverage ($\sim$ 85\%) for the 3CR sample by using
measurements from the 2MASS \citep{skrutskie06} or, when this is not
available, from HST images \citep{donzelli07}. For the BLO we also corrected
the host luminosity for the contribution of their bright IR nuclei, measured
by \citet{baldi09}.  In order to compare our results with those of
\citet{ledlow96} we used the color correction from \citet{mannucci01}, R - H =
2.5, and scaled the 1.4 GHz data to 178 MHz adopting a radio spectrum in the
form $F_{\nu} \propto \nu^{-0.7}$.

The relative scarcity of FR~I sources in the 3CR sample prevents us from
exploring in detail the host magnitude-dependent separation between the FR
classes.  However, the FR location for our sample is consistent with the
separation introduced by \citet{ledlow96}. We also checked that this result
holds using radio luminosities at 1.4 GHz as well as host magnitude in other
bands (i.e. V band). In line with their results we find a few exceptions,
associated with FR~I sources of extremely high radio powers in very massive
hosts.

In the right panel we introduced the optical spectroscopic classification,
separating the 3CR sources into HEG and LEG. For the LEG class we further
consider the FR type. HEG and LEG/FR~II sources are well mixed above the
FR~I/FR~II separation, having the same median in terms of radio power, and
only a small offset in the median host magnitude, with LEG $\sim$ 0.3 mag
brighter. As originally suggested by \citet[ hereafter GC01]{ghisellini01b}
the radio - host galaxy luminosity plane can be translated into a jet-power vs
black hole mass diagram. The substantial superposition between HEG and LEG
indicates that also the distribution of the estimated black hole masses and
jet power for the two classes are similar.

\begin{figure}[tbp]
  \centerline{ 
    \psfig{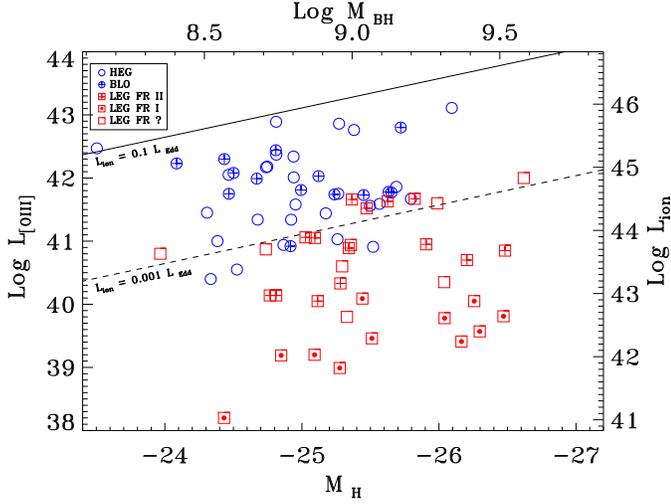}}
  \caption{ \label{ledlow3} Comparison of [O~III] line luminosity (in erg
    s$^{-1}$) and H band host magnitude for 3CR sources.  Symbols are as in
    the right panel of Fig. \ref{ledlow}, i.e.  HEG = blue circles, LEG = red
    squares (filled for FR~II, crossed for FR~I, empty for uncertain FR type.
    The right axis reports the ionizing luminosity (in erg s$^{-1}$) while the
    upper axis reports the estimated black hole mass in solar units.  The
    solid and dashed lines correspond to $L_{\rm ion} = 0.1 \cdot \, {\rm
      L_{Edd}}$ and $0.001 \cdot \, {\rm L_{Edd}}$, respectively.} 
 \end{figure}

 \citetalias{ghisellini01b} also noted that the line luminosity can be used as
 indicator of the ionizing luminosity, $L_{\rm ion}$, that they estimated from
 the radio power \citep[see e.g.][]{willott99}.  We can now derive this
 quantity directly from the [O~III] luminosity as\footnote{ Following
   \citet{rawlings91} the ionizing radiation is related to the NLR total
   luminosity as $L_{\rm ion} \sim $ C $L_{\rm NLR}$, where C is the covering
   factor of the NLR, for which we assume a value of 0.01.  $L_{\rm NLR}$ is
   derived as $L_{\rm NLR} = 3 \cdot (3 L_{\rm [O~II]} + 1.5 L_{\rm [O~III]})$
   and $L_{\rm [O~III]} = 4 L_{\rm [O~II]}$.}
$$ {\rm log \,L}_{{\rm ion}} \sim
{\rm log \,L}_{\rm{[O~III]}} + 2.83. $$
We also derived the black hole mass
from the host luminosity using the 
correlation of \citet{marconi03} 
$$
{\rm log (M}_{\rm BH}/{\rm M}_{\odot}) = -2.77 -0.464 {\rm M_H}.$$ In Fig.
\ref{ledlow3} we show the $L_{\rm [O~III]}-M_{\rm H}$ plane that, as explained
above, can be translated into a $L_{\rm ion}-M_{\rm BH}$ plane. With respect
to Fig. \ref{ledlow}, the location of LEG and HEG are substantially offset.
LEG/FR~I are located at the bottom of the diagram and all have $L_{\rm
  ion}$/$L_{\rm Edd}\lesssim10^{-4}$, where $L_{\rm Edd}= 1.3 \cdot 10^{38}
M_{\rm BH}$/$M_{\odot}$ erg s$^{-1}$ is the Eddington luminosity. LEG/FR~II
(and also LEG of uncertain FR type) lie in an intermediate area, while HEG
cover the top region. In general LEG (regardless of the FR type) and HEG are
rather well separated, but they coexist over the range $4 \times 10^{-4}
\lesssim L_{\rm ion}$/$L_{\rm Edd} \lesssim 2 \times 10^{-3}$. The dividing
line has apparently a positive slope, in the sense that LEG can be found at
high values of $L_{\rm[O III]}$ only in the brightest galaxies.

\citet{marchesini04} found that the bolometric luminosity for radio-loud AGN
has a bimodal distribution, related to a different level of $L_{\rm bol}$ for
LEG and FR~I with respect to BLO and radio-loud QSO, while, restricting only 
to RG, the evidence for bimodality was marginal.
We find a similar marginal result ($P \sim 0.1$) 
considering the ionizing luminosity or
the $L_{\rm ion}$/$L_{\rm Edd}$ ratio.

\section{Discussion} 
\label{discussion} 
\subsection{On the origin of the spectroscopic  sub-populations in radio-loud AGN} 
\label{subpop} 

The main result derived
from the spectroscopic properties of the 3CR sources is the presence of two
dominant populations of galaxies, HEG and LEG, separated on the basis
of the ratio between the intensity of high and low excitation 
emission lines. This finding confirms the 
original suggestion by \citet{laing94}. 

HEG and LEG differ also from the point of view of other properties: 1) we find
prominent broad lines in $\sim$30\% of HEG, but not in LEG; 2) HEG are
brighter than LEG in the [O~III] line by a factor of $\sim$ 10 at the same
total radio luminosity; 3) the two classes are well offset in a plane that
compares their ionizing luminosity and black hole mass; 4) the luminosity of
the nuclear sources in LEG is on average $\sim$ 30 times fainter than in HEG
(considering only HEG with broad lines, where the nucleus is not affected by
obscuration) at both optical and infrared wavelengths
\citep{chiaberge:fr2,baldi09} at a given radio core power; 5) HEG hosts
  are bluer than those of LEG, 
an indication that star formation is commonly
  associated with HEG while it rarely seen in LEG \citep{baldi08,smolcic09}.

Conversely, the distribution of black hole masses for the two
classes is not largely different. Actually, the overall black hole mass
distribution of these RG is rather narrow, with the vast majority
of the objects confined in the range 8.5 $\lesssim$ log $M_{\rm
  BH}$/$M_{\odot} \lesssim 9.5$ \citep[see also ][]{marchesini04}. Thus, HEG
nuclei are associated with a higher rate of radiative emission with respect to
LEG, not only in absolute values, but also in terms of fraction of the
Eddington level. 

\citetalias{ghisellini01b} suggested that the transition from FR~I to FR~II
can be associated with a threshold in the accretion rate that corresponds to a
change in the properties of the accretion disk, from ``standard'' to low
radiative efficiency (hereafter RIAF). \citet{nagao02} argued that the
differences in terms of emission line ratios between high and low excitation
sources can be ascribed to a change in the Spectral Energy Distribution of the
active nucleus, due to such a transition. In low efficiency disks, the
emerging spectrum is presumably harder than for a radiatively efficient disk.
This favors the emission of lower excitation lines, produced in the region of
partially ionized gas, excited by the higher energy photons. This is 
consistent with the observed spectroscopic differences between HEG and LEG. 
Additional support for this conclusion is provided if we 
consider radio-quiet AGN.  The
photoionization models derived by \citetalias{kewley06b} indicate that
Seyferts require a higher ionization parameter than LINERs, but the optical
line ratios of the more powerful objects of this latter class are better
reproduced assuming also a harder nuclear continuum.

Nevertheless, the location of the various classes in the $L_{\rm ion}$-$M_{\rm
  BH}$ plane is not straightforward to interpret in this scenario. In fact, we
lack of clear indications on how to convert the ionizing luminosity into
accretion rate for the (putative) RIAF disks. The separation between FR~I/LEG
and FR~II/HEG is sufficiently large (a factor of 100 - 1000 in $L_{\rm
  ion}$/$L_{\rm Edd}$) that one can probably infer that the latter class has
not only a higher ionizing luminosity but also a higher accretion rate.
Conversely, the brightest LEG overlap with FR~II/HEG; in the case they were
associated with RIAF, the superposition in terms of accretion rate would
become even larger, due to their lower radiative efficiency.

Furthermore, LEG span the whole range of radio power covered by the subsample
of 3CR sources with z $<$ 0.3 and, actually, the brightest radio source of our
sample (3C~123) is a LEG. Despite our rather strict criteria for the
definition of the FR type, we found a substantial fraction (16 out of 37) of
LEG with a FR~II morphology. Moreover the radio core power of LEG is similar
to that measured in HEG; restricting to the range of extended radio power 33
$<$ log $L_{178} <$ 34.5, where we have the bulk of the FR~II population, the
median core dominance is log $P_{\rm core}$ / $L_{178} = -2.69 \pm$ 0.13 in
LEG and log $P_{\rm core}$ / $L_{178} = -2.99 \pm$ 0.14 in HEG.  HEG and
LEG/FR~II are thus essentially indistinguishable from the point of view of
their radio properties, suggesting that the two classes share the same range
also in terms of jet power.

If the change in the spectral and nuclear properties from LEG to HEG are
associated with a threshold in the accretion rate, the similarity of the radio
properties of the two sub-populations requires a process of jet launching
essentially decoupled from the level of accretion. This is the case, for
example, in the \citet{blandford77} process where jets are shaped by the
magnetic field and the black hole spin and not univocally by the accretion
mechanism.

Focusing purely on those galaxies with FR~II morphology, an alternative
possibility is that the link between HEG and LEG is related to variability can
be excluded. In fact, in response to a decrease in the accretion rate, the
luminosity of the optical nucleus, of the BLR, and of the NLR, as well as its
excitation level, will decline, possibly causing a change from a HEG to a LEG.
However, this contrasts with the requirement that the radio core emission must
remain essentially unaltered while this component reacts on very short
timescales to changes of the central activity.

Based on the substantial difficulties in accounting for the similarities and
differences between LEG and HEG based on different or time-varying accretion
rates, we here propose a different interpretation for the origin of the
spectroscopic sub-populations of radio-loud AGN.  We speculate that the
separation between LEG and HEG is not due to a different rate of accretion
but, instead, to a different {\sl mode} of accretion. In this scenario, HEG
are powered by accretion of cold gas, e.g. provided by a recent merger with a
gas rich galaxy \citep[e.g. ][]{baldi08}.  Cold gas, approaching the central
regions of the galaxy, forms the various structures commonly seen in these
AGN, such as a molecular torus, a Broad Line Region, and a standard,
geometrically thin, accretion disk. Conversely, LEG accrete hot material,
provided by the ample reservoir of their X-ray emitting gaseous coronae. This
process has been shown to be able to account for the nuclear activity of FR~I
radio-galaxies \citep{allen06,balmaverde08} extending up to a radio power of
log $L_{178} \sim 33$. The temperature of the accreting gas is typically
around 1 keV, $\sim 10^7$ K.  This prevents the formation of the ``cold''
structures, in particular of a molecular torus, but also of the clouds of the
BLR, whose ionized portion has a temperature of $\sim 10^5$ K, unless the
in-flowing gas is able to cool dramatically on its way to the center of the
galaxy. Indeed BLR are not seen in LEG and they also do not generally show the
high level of absorption in the X-ray band expected if their nuclei were seen
through an obscuring torus (e.g. \citealt{hardcastle09}).

Moreover, the properties of the accretion disk are likely to be substantially
altered due to the high initial temperature of the gas. From a geometrical
point of view, a ``cold'', standard accretion disk is flattened by rotation,
while it remains geometrically thick for higher temperatures.  While the
radiative emission in a standard disk is dominated by UV and soft X-ray
photons, a hotter disk emits most of the radiation at higher energies.  The
higher temperature in case of hot accretion corresponds also to a lower
radiative efficiency due to the reduced gas cooling at these temperatures (but
see \citealt{dumont92}, for the effects of high density on the gas cooling
function). At a given accretion rate, the number of ionizing photons is then
reduced, due to the combination of higher average photon energy and lower
overall emission. The emitted spectrum is harder and it produces lines of
lower excitation, as discussed above. This effect can produce the spectral
separation between LEG and HEG similarly to RIAF.

Furthermore, \citet{capetti:cccriga} showed that the non-thermal cores in FR~I
produce a sufficient flux of high energy photons to account for the ionization
of their NLR. The radio and optical luminosities of the nuclei of FR~I and
FR~II/LEG are linked by a common linear correlation \citep{chiaberge:fr2} and,
as shown by Fig. \ref{o3re}, this applies also to radio cores and line
luminosity. It is then possible that the dominant source of ionizing photons
in LEG must be ascribed to non-thermal emission, associated with the
base of their jets, rather then with their accretion disks.

The accretion rate in FR~II/LEG can be estimated by extrapolating the scaling
relation between jet and accretion power derived for low luminosity FR~I/LEG
to the objects of highest luminosity of this class.  \citet{balmaverde08}
estimated that the accretion rate needed to power a radio source with $L_{178}
\sim 10^{33}$ erg s$^{-1}$ is $P_{\rm accr} \sim 10^{44.6}$ erg s$^{-1}$.
This requires $P_{\rm accr} \sim 10^{46}$ erg s$^{-1}$ for the most powerful
LEG, corresponding to a fraction $\dot{\rm m} \sim 0.1$ of the Eddington rate
for a 10$^9 {\rm M}_{\sun}$ black hole. These hot flows at high accretion rate
can be probably associated with the optically thin, geometrically thick
solutions (e.g. \citealt{abramowicz95}) that can reach, for a viscosity
parameter $\alpha =0.1$, a rate of the order of the Eddington one.

The mechanism of jet launching, likely to be determined by the disk structure
in the region closer to the black hole, might not be sensitive to the gas
history, but only to the final accretion rate. Provided that this reaches
comparable high levels in HEG and LEG, radio structures of similar morphology
and power can be formed in the two sub-classes.

\subsection{Radio-loud vs radio-quiet (SDSS) AGN}
\label{rqrl}

The separation between HEG and LEG is reminiscent of that found by
\citetalias{kewley06b} for the SDSS sources, mostly radio-quiet AGN. However,
we find a significant number of LEG located above the line marking the
transition between LINERs and Seyferts.  The location of LEG shows an upward
scatter with respect to the `finger' of highest LINERs density by $\sim$0.2
dex in the [O III]/\Hb\ ratio.  As already noted there is a substantial
mismatch in luminosity between the 3CR and the SDSS sources. Our data are not
sufficient to conclude whether this is due to a genuine difference between the
(mostly) radio-quiet AGN of the SDSS and the radio-loud AGN of the 3CR sample,
(due e.g. to a contribution of jets emission to the line excitation) or simply
to a luminosity difference.

Similarly, we noted that the 3CR sources are concentrated along the edges of
the SDSS density distribution. The first possibility to account for the
location of SDSS and 3CR sources is again a general difference in the
spectroscopic behavior between radio-quiet and radio-loud AGN. Alternatively,
the location of the SDSS sources could be due to the contamination of star
forming regions. In this scenario, this results in a large spread in the line
ratios, depending on the relative contribution of the line emission produced
by the active nucleus and by star formation. The sources would be then
distributed along a mixing region (possibly the `fingers' seen in Figs.
\ref{dd} and \ref{sdss4045}), ranging from a `pure' AGN to a star forming
spectrum. In the case of 3CR sources, the higher line luminosity is likely to
be indication of a dominant AGN contribution, also considering the general
weakness of star formation in their early type host with respect to later
types.

The analysis of radio-loud AGN of lower luminosity can be used to
test these alternative hypothesis.

\section{Summary and conclusions} 
\label{fine}

We used measurements of emission lines for a 92 \% complete survey of the 3CR
sample of radio-galaxies with z $<$ 0.3 to explore their spectroscopic
properties. 
The different steps used to classify the 3CR 
sources in the various sub-populations can be summarized as follows:

$\bullet$ {\it Excitation Index diagram:} we first considered a new
spectroscopic indicator, 
the Excitation Index, defined as 
E.I. = log [O~III]$/$\Hb - 1/3 (log
[N~II]$/$\Ha + log [S~II]$/$\Ha + log [O~I]$/$\Ha) and representing the
relative intensity of high and low excitation lines. The presence of a bimodal
distribution of E.I. allows us to define two main spectroscopic classes, LEG
and HEG, with LEG being sources with E.I.
$\lesssim$ 0.95.  The drawback of this definition is that it requires the
measurements of all 6 diagnostic lines.

$\bullet$ {\it Diagnostic diagrams:} the close correspondence between the
excitation index and the location in the 3 `standard' diagnostic diagrams
enabled us to define a class membership in a less ideal situation, i.e. when
in addition to \Ha, \Hb\ and [O III], only one or two lines among [S II], [O
I] or [N II] can be measured, defining as LEG all sources for which:

log [O~III]/\Hb\ - log  [N~II]/\Ha\ $\lesssim$ 0.7,

log [O~III]/\Hb\ - log  [S~II]/\Ha\  $\lesssim$ 0.9, or

log [O~III]/\Hb\ - log  [O~I]/\Ha\ $\lesssim$ 1.4.

$\bullet$ {\it Optical line-radio correlation.} We also used the $L_{\rm
  [OIII]}$ vs $L_{178 MHz}$ correlation to extend the classification for
objects with the only requirement of a [O~III] line measurement.  This is due
to the rather well defined separation in line emission, at fixed radio power,
between HEG and LEG. In particular we found

$-1 \lesssim {\rm log} ({L_{\rm [O III]}}/\nu L_{178}) \lesssim  \,\, 0.5$ for HEG

$-2 \lesssim {\rm log} ({L_{\rm [O III]}}/\nu L_{178}) \lesssim -0.5$ for LEG.

Using this strategy we were able to associate a spectroscopic class to 87
sources, representing 84 \% of the sample. The breakdown in the various
sub-classes is of 46 HEG (16 of which are BLO) and 37 LEG.
In addition to the two main sub-classes, we found one object with a spectrum
typical of star-forming galaxies, and
3 sources of extremely low excitation (ELEG). 17 galaxies remain
spectroscopically unclassified.  None of the unclassified sources is
consistent with being a HEG, as they all have very low values of $L_{\rm
  [OIII]}/\nu L_{178}$.  However, we cannot conclude in general that they are
LEG, since they could belong to the ELEG class. 

The ELEG are particularly intriguing, being well offset from the rest of 3CR
sample in all diagnostic diagrams. They are also characterized by very low
values of the ratio between [O~III] and radio luminosity as well as of core
dominance. We defer to a forthcoming paper a more detailed analysis of these
sources.

The dual population of HEG and LEG is reminiscent of the Seyfert and LINERs
subclasses found in the SDSS emission line galaxies. Indeed, at zero-th order,
the spectroscopic properties of radio-loud AGN of the 3CR and the (mostly)
radio-quiet AGN of the SDSS are rather similar. However, looking at finer
details, the separation between LEG and HEG shows an upward
scatter by $\sim$0.2 dex in the [O III]/\Hb\ ratio with respect to the
transition from LINER to Seyfert. Furthermore, the 3CR sources are
concentrated along the edges of the SDSS density distribution in the
spectroscopic diagnostic diagrams. Both results could be related to the
substantial mismatch in luminosity between the two samples, the 3CR being
brighter by an average factor of 30 in emission line than the SDSS sources.
However, we cannot exclude that the location of SDSS and 3CR source is an
indication of a genuine difference in the spectroscopic behavior of
radio-quiet and radio-loud AGN. The analysis of radio-loud AGN of lower
luminosity can be used to test these alternative hypothesis.

Let us now summarize the similarities and differences between the two main
spectroscopic classes of RG.  All broad-line objects are HEG from
the point of view of their narrow emission line ratios, but no LEG shows broad
line in its spectrum.  While all HEG are associated with FR~II radio-galaxies,
LEG are of both FR~I and FR~II type.  HEG are only found in relatively bright
radio sources, log $L_{178} \gtrsim 32.8$. Instead LEG
cover the whole range of radio power (30.7 $\lesssim$ log $L_{178} \lesssim$
35.4), the brightest of them being FR~II.  HEG and LEG obey to two linear
correlations between line and radio emission, with a slope consistent with
unity and a rms $\lesssim 0.5$ dex. HEG are brighter than LEG by a factor of
$\sim 10$ in [O~III] line.

Conversely, the substantial superposition of the host galaxy luminosity of the
two classes suggests that the distribution of their black hole masses is not
strongly different, with the vast majority of the objects confined in the
range 8.5 $\lesssim$ log $M_{\rm BH}$ / $M_{\odot} \lesssim 9.5$. Considering
only the LEG with a FR~II morphology, a further similarity with HEG emerges
from the analysis of their radio structure. LEG/FR~II extend over the same
range of radio power of HEG and have the same level of radio core dominance.
The fact that the radio properties of LEG/FR~II and HEG are essentially
indistinguishable is an indication that the two classes share the same range
of jet power.

While the differences in terms of line intensities and line ratios
(as well as in optical and IR nuclear luminosities) can be 
interpreted in a framework in which LEG are
associated with a lower accretion rate than HEG, the similarity of the radio 
properties of the two sub-populations (having established that they do not
differ in terms of black hole mass) would require that the process of jet 
launching is essentially decoupled from the level of accretion.
This is the case, for example, of the Blandford-Znajek process
that envisages jets powered by the extraction 
of rotational energy from a spinning black hole.

We instead speculated that the separation between LEG and HEG is due to a
different mode of accretion: while HEG are powered by cold gas, LEG accrete
hot gas, a mechanism that has been already demonstrated to be able to account
for the activity of FR~I radio-galaxies.  The high initial temperature of the
in-flowing gas in LEG prevents the formation of the various ``cold'' AGN
structures, such as molecular tori and Broad Line Regions, seen in HEG but not
in LEG.  The radiative output of a ``hot'' accretion disk is reduced with
respect to a standard geometrically thin disk, at a given accretion rate, due
to the lower cooling at these temperatures and it will be emitted mostly at
high energies. The harder spectrum and the lower number of ionizing photons
gives rise to emission lines of lower luminosity and excitation, producing the
spectral separation in lines properties between LEG and HEG. Since the
mechanism of jet launching operates in the innermost regions of the accretion
flow, it is likely that the memory of the initial gas conditions are
effectively canceled and that the jet power is sensitive only to the final
accretion rate.

The attractiveness of this scheme is that it can reproduce simultaneously, by
only varying the initial temperature of the accreting gas, the various
differences between LEG and HEG. Clearly this hypothesis must be analyzed in
greater detail, exploring the properties of the proposed hot accretion flows
combined to a high accretion rate. In particular a crucial requirement is that
the coronal accreting gas must be able to remain hot during the inflow. The
proposed scenario can also be tested and constrained using the limits on the
nuclear LEG emission in the various observing bands. It is also of great
interest to establish whether LEG can be found in objects of at most moderate
redshift (and, consequently, power) such as those considered here, or,
conversely, can be associated even with radio galaxies of the highest
luminosity.

Summarizing, we found that the radio-galaxies in the 3CR sample are
  composed by two main populations, separated on the basis of their emission
  line properties. However, this does not correspond simply to a 
  difference in their radio properties. We propose a scenario in which the
  two classes are associated with a different mode of accretion, set by the
  initial temperature of the inflowing gas.

{\sl Acknowledgments.} 
SB and ACe acknowledge the Italian MIUR for financial
support.
This publication makes use of data products from the Two Micron All Sky
Survey, which is a joint project of the University of Massachusetts and the
Infrared Processing and Analysis Center/California Institute of Technology,
funded by the National Aeronautics and Space Administration and the National
Science Foundation.

\bibliographystyle{aa}

\end{document}